\documentstyle{mn}

\newif\ifAMStwofonts

\ifoldfss
  \ifCUPmtlplainloaded \else
    \NewTextAlphabet{textbfit} {cmbxti10} {}
    \NewTextAlphabet{textbfss} {cmssbx10} {}
    \NewMathAlphabet{mathbfit} {cmbxti10} {} 
    \NewMathAlphabet{mathbfss} {cmssbx10} {} 
  \fi
  \ifAMStwofonts
    \ifCUPmtlplainloaded \else
      \NewSymbolFont{upmath} {eurm10}
      \NewSymbolFont{AMSa} {msam10}
      \NewMathSymbol{\upi}     {0}{upmath}{19}
      \NewMathSymbol{\umu}     {0}{upmath}{16}
      \NewMathSymbol{\upartial}{0}{upmath}{40}
      \NewMathSymbol{\leqslant}{3}{AMSa}{36}
      \NewMathSymbol{\geqslant}{3}{AMSa}{3E}

      \let\leq=\leqslant 
      \let\geq=\geqslant 
    \fi
  \fi
\fi 

\ifnfssone
  \newmathalphabet{\mathit}
  \addtoversion{normal}{\mathit}{cmr}{m}{it}
  \addtoversion{bold}{\mathit}{cmr}{bx}{it}
  \newmathalphabet{\mathbfit} 
  \addtoversion{normal}{\mathbfit}{cmr}{bx}{it}
  \addtoversion{bold}{\mathbfit}{cmr}{bx}{it}
  \newmathalphabet{\mathbfss} 
  \addtoversion{normal}{\mathbfss}{cmss}{bx}{n}
  \addtoversion{bold}{\mathbfss}{cmss}{bx}{n}
  \ifAMStwofonts
    \ifCUPmtlplainloaded \else
      \UseAMStwoboldmath
      \makeatletter
      \new@mathgroup\upmath@group
      \define@mathgroup\mv@normal\upmath@group{eur}{m}{n}
      \define@mathgroup\mv@bold\upmath@group{eur}{b}{n}
      \edef\UPM{\hexnumber\upmath@group}
      \new@mathgroup\amsa@group
      \define@mathgroup\mv@normal\amsa@group{msa}{m}{n}
      \define@mathgroup\mv@bold\amsa@group{msa}{m}{n}
      \edef\AMSa{\hexnumber\amsa@group}
      \makeatother
      \mathchardef\upi="0\UPM19
      \mathchardef\umu="0\UPM16
      \mathchardef\upartial="0\UPM40
      \mathchardef\leqslant="3\AMSa36
      \mathchardef\geqslant="3\AMSa3E

      \let\leq=\leqslant 
      \let\geq=\geqslant 
    \fi
  \fi
\fi 

\ifnfsstwo
  \DeclareMathAlphabet{\mathbfit}{OT1}{cmr}{bx}{it}
  \SetMathAlphabet\mathbfit{bold}{OT1}{cmr}{bx}{it}
  \DeclareMathAlphabet{\mathbfss}{OT1}{cmss}{bx}{n}
  \SetMathAlphabet\mathbfss{bold}{OT1}{cmss}{bx}{n}
  \ifAMStwofonts
    \ifCUPmtlplainloaded \else
      \DeclareSymbolFont{UPM}{U}{eur}{m}{n}
      \SetSymbolFont{UPM}{bold}{U}{eur}{b}{n}
      \DeclareSymbolFont{AMSa}{U}{msa}{m}{n}
      \DeclareMathSymbol{\upi}{0}{UPM}{"19}
      \DeclareMathSymbol{\umu}{0}{UPM}{"16}
      \DeclareMathSymbol{\upartial}{0}{UPM}{"40}
      \DeclareMathSymbol{\leqslant}{3}{AMSa}{"36}
      \DeclareMathSymbol{\geqslant}{3}{AMSa}{"3E}

      \let\leq=\leqslant 
      \let\geq=\geqslant 
    \fi
  \fi
\fi 

\ifCUPmtlplainloaded \else
  \ifAMStwofonts \else 
    \def\upi{\pi}
    \def\umu{\mu}
    \def\upartial{\partial}
  \fi
\fi

\title[A NICMOS imaging study of quasar hosts]{A NICMOS imaging study 
of high-z quasar host galaxies}
\author[M. J. Kukula et al.]
       {Marek J. Kukula,$^{1}$\thanks{M.Kukula@roe.ac.uk} James
        S. Dunlop,$^{1}$ Ross J. McLure,$^{2}$ Lance Miller,$^{2}$
\newauthor
        Will J. Percival,$^{1}$ Stefi A. Baum$^{3}$ and Christopher P. O'Dea,$^{3}$ \\
        $^{1}$Institute for Astronomy, University of Edinburgh, Royal Observatory, Edinburgh EH9 3HJ, U.K. \\
        $^{2}$Nuclear and Astrophysics Laboratory, University of Oxford,
Keble Road, Oxford, OX1 3RH, U.K. \\
        $^{3}$Space Telescope Science Institute, 3700 San Martin Drive,
Baltimore, MD 21218, U.S.A.}

\date{  }

\pagerange{\pageref{firstpage}--\pageref{lastpage}}
\pubyear{2001}

\begin{document}

\maketitle

\label{firstpage}

\begin{abstract}
We present the first results from a major Hubble Space Telescope
programme designed to investigate the cosmological evolution of quasar
host galaxies from $z \simeq 2$ to the present day. Here we describe
$J$ and $H$-band NICMOS imaging of two quasar samples at redshifts of
0.9 and 1.9 respectively. Each sample contains equal numbers of
radio-loud and radio-quiet quasars, selected to lie within the same
narrow range of optical absolute magnitude ($-24\geq M_{V} \geq
-25$).  Filter and target selection were designed to ensure that at
each redshift the images sample the same part of the object's
rest-frame spectrum, longwards of 4000\AA~ where starlight from the
host galaxy is relatively prominent, but avoiding potential
contamination by [O{\sc iii}]$\lambda5007$ and H$\alpha$ emission
lines.

At $z\simeq1$ we have been able to establish host-galaxy luminosities
and scalelengths with sufficient accuracy to demonstrate that the
hosts of both radio-loud and radio-quiet quasars lie on the same
Kormendy relation described by 3CR radio galaxies at comparable
redshift (McLure \& Dunlop 2000).  Taken at face value the gap
between the host luminosities of radio-loud and radio-quiet objects
appears to have widened from only $\simeq 0.4$ mag. at $z \simeq 0.2$
(Dunlop et al. 2001) to $\simeq 1$ mag. at $z \simeq 1$, a difference
that cannot be due to emission-line contamination given the design of
our study. However, within current uncertainties, simple passive
stellar evolution is sufficient to link these galaxies with the
elliptical hosts of low-redshift quasars of comparable nuclear output,
implying that the hosts are virtually fully assembled by $z\sim1$.

At $z\simeq2$ the hosts have proved harder to characterise accurately,
and for only two of the nine $z \simeq 2$ quasars observed has it
proved possible to properly constrain the scalelength of the host
galaxy. However, the data are of sufficient quality to yield
host-galaxy luminosities accurate to within a factor $\simeq 2$. At
this redshift the luminosity gap between radio-loud and radio-quiet
quasars appears to have widened further to $\simeq 1.5$ mag. Thus
while the hosts of radio-loud quasars remain consistent with a
formation epoch of $z > 3$, allowing for passive evolution implies
that the hosts of radio-quiet quasars are $\simeq 2-4$ times less
massive at $z \simeq 2$ than at $z \simeq 0.2$.

If the relationship between black-hole and spheroid mass is unchanged
out to redshift $z \simeq 2$, then our results rule out any model of
quasar evolution which involves a substantial component of luminosity
evolution (e.g. Kauffmann \& Haehnelt 2000). Rather, this study
indicates that at $z \simeq 2$ there is a substantial increase in the
number density of active black holes, along with a moderate increase
in the fueling efficiency of a typical observed quasar. The fact that
this latter effect is not displayed by the radio-loud objects in our
sample might be explained by a selection effect arising from the fact
that powerful radio sources are only produced by the most massive
black holes (Dunlop et al. 2001; McLure \& Dunlop 2000b).
  
\end{abstract}


\section{Introduction}

Recent years have brought great advances in our understanding of the
symbiotic relationship between active galactic nuclei (AGN) and the
galaxies in which they occur. Some of the most impressive advances
concern quasars - the most powerful known AGN -  as improvements in
ground-based observing techniques and the advent of the Hubble Space
Telescope (HST) have allowed the diffuse `fuzz' of the underlying host
galaxy to be reliably separated from the wings of the bright non-stellar
nuclear point spread function (PSF) for the first time.

The problems inherent in observing quasar host galaxies from the
ground are too well-known to require detailed explanation but,
considering the enormous difficulties involved in separating faint,
diffuse galaxy light from the PSF of a bright quasar, it is perhaps
surprising that so much progress has been made to date using
ground-based techniques. Although such studies are effectively limited
to $z \leq 0.3$, at these low redshifts a combination of ground-based
and HST programmes is beginning to yield a coherent picture
of the properties of quasar hosts in the local universe.

However, the local universe is not the most representative region in
which to study the quasar population: arguably the epoch of greatest
importance to quasar research occurred at redshifts of $z\sim2$ to 3,
when quasars were $2 - 3$ orders of magnitude more numerous than they are
today. Although simulations show that it is very difficult to derive the
properties of quasar hosts at such high $z$ from the ground with any
degree of confidence (Taylor et al. 1996), numerous attempts have been
made with (not surprisingly) confusing results. In this paper we
present the results of a study using the Near-Infrared Camera and
Multi-Object Spectrometer (NICMOS) on HST to investigate how the
luminosities, sizes and morphologies of the hosts of both radio-loud
and radio-quiet quasars (RLQs \& RQQs) have evolved from the `golden
era' of quasar activity to the present day.

\subsection{The host galaxies of low-redshift quasars}

At low redshifts, attention has been focused on determining the
luminosities, scalelengths, morphologies and interaction histories of
quasar host galaxies, and investigating the extent to which these
properties are correlated to the optical and radio luminosity of the
quasar (V\'{e}ron-Cetty \& Woltjer 1990; Dunlop et al. 1993; McLeod \&
Rieke 1994a,b; Bahcall, Kirhakos \& Schneider 1995a,b,1996; Hutchings
\& Morris 1995; Disney et al. 1995; Taylor et al. 1996; Bahcall et
al. 1997; Hooper, Impey \& Foltz 1997; Boyce et al. 1998; Carballo et
al. 1998; McLeod, Rieke \& Storrie-Lombardi 1999; Schade, Boyle \&
Letawsky 2000; Hamilton, Casertano \& Turnshek 2001; M\'{a}rquez et
al. 2001).  For example, it has long been known that low-luminosity
AGN display marked preferences in terms of host type, with
(radio-quiet) Seyferts favouring spiral hosts whilst (radio-loud)
Radio Galaxies are exclusively associated with massive
ellipticals. However, recent studies have demonstrated that this
distinction breaks down at the higher redshifts and nuclear
luminosities typical of radio-loud and radio-quiet quasars, and that
powerful nuclear activity is predominantly associated with
bulge-dominated galaxies, regardless of radio luminosity.

In our own HST study of low redshift ($0.1\leq z\leq0.25$), low
luminosity ($-23\leq M-{V} \leq -26$) RLQs and RQQs (McLure et
al. 1999; Dunlop et al. 2001) we found that for quasars brighter than
$M_{R}\sim -24$ the hosts are invariably massive elliptical galaxies
with $L\geq 2L^{\star}$, with typical scalelengths of $\sim 10$~kpc, and
display a Kormendy relation identical to that of brightest cluster
galaxies. This result is consistent with measurements of the massive
dark objects in the nuclei of nearby inactive galaxies, which predict
that only the largest spheroidal systems will harbour black holes of
the requisite mass to produce luminous quasars (Kormendy \& Richstone
1995; Magorrian et al. 1998; van der Marel 1999), and also with the
discovery by McLeod \& Rieke (1995) of a lower limit to the mass of a
quasar host which is correlated with the luminosity of the quasar. The
implication of these studies is that massive elliptical galaxies are
the parent population of the quasar phenomenon. However, in a
ground-based $K$-band study of luminous ($-25 \geq M_{V} \geq -27$)
RQQs Percival et al. (2001) found evidence that, on scales much larger
than those probed in optical HST images, some of the hosts are
dominated by a disc component. Clearly the issue of host morphology
has yet to be entirely resolved, but the finding that a quasar host
requires at the very least a massive spheroidal component seems
secure.

Spectral energy distributions (SEDs) also provide important
information on the nature of the host galaxies.  McLure et al. (1999)
and Dunlop et al. (2001) found that the $R-K$ colours of their quasar
hosts were consistent with normal passively-evolving elliptical
galaxies with ages of $\sim 12$~Gyr, implying that the stellar
populations formed at high redshift ($z\sim3$). Optical off-nuclear
spectroscopy of the same objects (Hughes et al. 2000; Nolan et
al. 2001) confirms that, longwards of the 4000\AA~break in their
restframe spectra, these hosts are dominated by light from an old,
well-established population of stars. Similar results are obtained for
radio galaxies (e.g. de Vries et al. 2000).

\subsection{High-redshift quasar hosts} 

The fact that the host galaxies of low-redshift quasars have mature
stellar populations and contain a massive bulge component implies
that, like local inactive elliptical galaxies, they are arguably
consistent with the products of successive merger events.  This raises
important questions about the potential redshift dependence of the
properties of quasar hosts.  The dramatic cosmological evolution
exhibited by the quasar population itself, with a comoving number
density which peaks at redshifts of 2-3 before undergoing a rapid
decline to its present low value, has been known for many years (e.g.
Boyle et al. 1988; Dunlop \& Peacock 1990; Warren, Hewett \& Osmer
1994) but the extent and form of any evolution affecting quasar hosts
remain as largely unknown quantities. The issue has been given renewed
urgency by continuing efforts to characterise the starformation
history of the universe, for which the redshift range of $z=2-3$ also
appears to correspond to an important epoch (Madau et al. 1996; Hughes
et al. 1998; Steidel et al. 1999).

It is therefore important to determine how the luminosities,
scalelengths, and the degree of morphological disturbance seen in
quasar host galaxies varies with cosmological epoch between the peak
of quasar activity at $z \simeq 2$ and the present day, and whether
the hosts of radio-loud and radio-quiet quasars differ in their
luminosity or morphological evolution.  However, the difficulties
involved in such studies are far more formidable than at low $z$ and
this is reflected in the confusion which surrounds the interpretation
of ground-based attempts to observe high-redshift quasar hosts.

Several groups have attempted such observations from the ground and
have succeeded in detecting extensions around high-redshift quasars
(Hintzen, Romanishin \& Valdes 1991; Heckman et al. 1991; Lehnert et
al. 1992; Aretxaga, Boyle \& Terlevich 1995; R\"{o}nnback et al. 1996;
Aretxaga, Terlevich \& Boyle 1998). These studies suggest that the
host galaxies of quasars at $z \simeq 2$ are $\sim 2.5$ to $3$ mag
brighter than those of low-redshift quasars, a result which is
consistent with common scenarios for elliptical galaxy
evolution. However, the usefulness of any such direct comparison is
confused by the fact that these high-redshift quasars are typically 5
magnitudes more luminous than the low-redshift objects studied to
date. In addition, the observational situation is somewhat complex
since other workers have failed to detect extended emission around
high-redshift RQQs (Lowenthal et al. 1995). Hutchings (1995) also
concludes that the hosts of RQQs are considerably fainter than those
of RLQs at high redshift, a result which could however simply indicate
that much of the light detected around high-redshift RLQs is not due
to stars, but rather to processes associated with the extreme radio
activity, as is found for high-$z$ radio galaxies (Tadhunter et
al. 1992), and for low-$z$ RLQs (Stockton \& MacKenty 1987). This
emphasises the importance of avoiding emission lines and sampling the
rest-frame spectrum longwards of the 4000\AA\ break.

The study described in this paper attempts to avoid the problems
outlined above by observing test samples of quasars over a range of
redshifts out to $z=2$ but limited to a narrow range of optical
absolute magnitude ($-24 \leq M_{V} \leq -25$). Careful choice of
filters ensures that our images always sample the same
emission-line-free region of the object's restframe spectrum, thus
avoiding any regions of extended [O{\sc iii}]$\lambda5007$ and
H$\alpha$ emission which may be associated with the active
nucleus. Thus, at each redshift we are always observing objects of
roughly the same intrinsic luminosity, and in the same region
(approximately $V$-band) of their rest-frame spectrum.

The layout of this paper is as follows. In section 2 we provide
details of how our quasar samples were selected, and how the HST
NICMOS observations were designed and implemented. The process of data
reduction is described in section 3, including the details of how we
chose to tackle the particular problems associated with producing
NICMOS images of sufficient quality for a study of this type. In
section 4 we explain how host-galaxy parameters were extracted from
the data via 2-D modelling, and summarize the basic results of the
image analysis. Then in section 5 we explore the implications of
combining these new results with the results of our Cycle 6 HST study
of quasar hosts at much lower redshift ($z \simeq 0.2$; McLure et
al. 1999; Dunlop et al. 2001), and discuss our findings in the context
of existing theoretical predictions. Finally, our results are
summarized in section 6.  For ease of comparison with existing
results/predictions we assume an Einstein-de Sitter cosmology with
$H_{0}=50$~km~s$^{-1}$Mpc$^{-1}$ throughout most of this
paper. However, in the penultimate section we focus on the
implications of our results within the currently-favoured flat
cosmological model with $\Omega_{m} = 0.3$, $\Omega_{\Lambda} = 0.7$ and
$H_{0}=65$~km~s$^{-1}$Mpc$^{-1}$.

\begin{figure*}
\vspace{7.0cm}
\centering
\setlength{\unitlength}{1mm}
\includegraphics{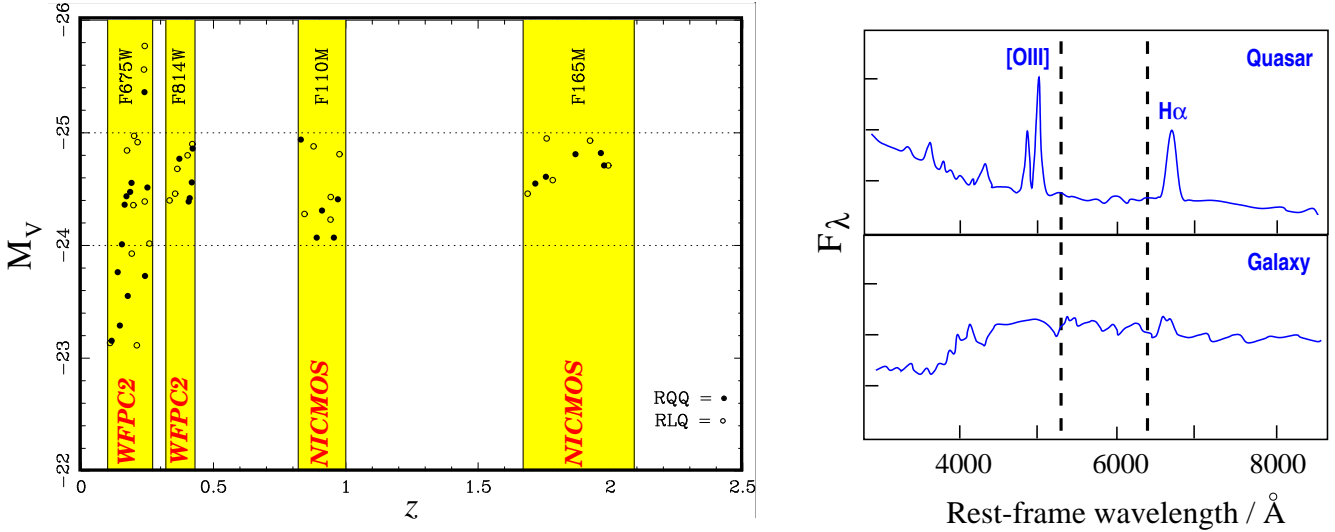}
\caption{Left: Absolute $V$ magnitude versus redshift for the RQQs
(filled circles) and RLQs (open circles) in the current study. The
shaded vertical bars represent the redshift regimes over which the
indicated WFPC2 and NICMOS filters remain free from contamination by
[O{\sc iii}]$\lambda5007$ and H$\alpha \lambda6563$ emission lines.
Also shown are the quasars from our previous HST imaging study using
WFPC2/F675W (McLure et al. 1999; Dunlop et al. 2001). This low-$z$
sample spans almost 3 magnitudes in optical luminosity at a given
redshift ($z \simeq 0.2$) whereas the three samples selected for the
current study are confined to only 1 mag in luminosity but span the
bulk of the history of the Universe.  Thus these complementary low-$z$
objects provide the baseline against which to measure any cosmological
evolution in the high-$z$ samples. Right: Schematic showing typical
rest-frame spectra of a quasar (upper panel) and an early-type galaxy
(lower panel). The two vertical lines delineate the region of the
host-galaxy spectrum which is targeted by the redshift/filter
combinations used in the observations described in this paper. Note
the avoidance of prominent quasar emission lines and the increase in
galaxy flux density longwards of the break feature at 4000\AA.}
\end{figure*}

\begin{table*}
 \centering
 \begin{minipage}{140mm}

 \caption{Summary of the quasar samples observed in the current NICMOS
 study. RQQs are derived from the UVX survey by Boyle et al. (1990)
 (SGP) and the survey of Marshall et al. (1984) (BVF), and are
 confirmed as radio quiet ($P_{5GHz}<10^{24.5}$W Hz$^{-1}$ sr$^{-1}$)
 by the VLA FIRST \& NVSS surveys. RLQs were selected from the
 V\'{e}ron-Cetty \& V\'{e}ron (1993) quasar catalogue, with reported
 5~GHz radio luminosities $P_{5GHz}>10^{25}$W Hz$^{-1}$ sr$^{-1}$ and
 steep radio spectra. $V$-band magnitudes are from various sources,
 and where necessary have been converted from $B$-band
 measurements. For the estimation of luminosities at both radio and
 optical wavelengths we have assumed a quasar spectrum of the form
 $f(\nu) \propto \nu^{-0.5}$}

\begin{tabular}{@{}lllllccl@{}}
\hline
            &       &\multicolumn{2}{c}{J2000 Position} &     &      &  Observing & Time on \\
 Name       &  Type & RA ($h~m~s$) & Dec $^{\circ}~'~''$& $z$ & $V$  &  date (dd/mm/yy)& source (sec) \\ \hline	  
 \multicolumn{8}{c}{$J$-band (F110M) sample ($z\sim1$)}\\ 
 SGP5:46    &  RQQ  & 00 52 22.76  & $-$27 30 02.8  & 0.955  & 18.88  & 25/08/97  &  $1\times2048$ \\
 SGP2:47    &  RQQ  & 00 53 02.24  & $-$29 12 55.5  & 0.830  & 17.17  & 14/08/97  &  $1\times2048$ \\
 BVF225     &  RQQ  & 13 04 10.47  & $+$35 36 50.8  & 0.910  & 18.54  & 16/02/98  &  $1\times2048$ \\
 BVF247     &  RQQ  & 13 05 05.04  & $+$35 51 20.6  & 0.890  & 18.73  & 15/02/98  &  $1\times2048$ \\
 BVF262     &  RQQ  & 13 05 30.91  & $+$35 17 13.5  & 0.970  & 18.57  & 18/02/98  &  $1\times2048$ \\
 PKS0440-00 &  RLQ  & 04 42 38.64  & $-$00 17 43.4  & 0.844  & 18.41  & 11/08/97  &  $1\times2048$ \\
 PKS0938+18 &  RLQ  & 09 41 23.17  & $+$18 21 06.0  & 0.943  & 18.49  & 18/02/98  &  $1\times2048$ \\
 3C422      &  RLQ  & 20 47 10.39  & $-$02 36 22.5  & 0.942  & 18.69  & 03/11/97  &  $1\times2048$ \\
 MC2112+172 &  RLQ  & 21 14 56.68  & $+$17 29 22.7  & 0.878  & 17.89  & 30/10/97  &  $1\times2048$ \\
 4C02.54    &  RLQ  & 22 09 32.82  & $+$02 18 40.9  & 0.976  & 18.19  & 05/11/97  &  $1\times2048$ \\
 SA107-626  &  PSF star & 15 40 05.31 & $-$00 17 29.2 & 0    & 12.77  & 30/07/97  &  $8\times256$  \\
 SA107-627  &  PSF star & 15 40 07.45 & $-$00 17 23.0 & 0    & 13.47  & 17/02/98  &  $8\times256$  \\ 
 \multicolumn{8}{c}{$H$-band (F165M) sample ($z\sim2$)}\\ 
 SGP2:36    &  RQQ   & 00 51 14.32  & $-$29 05 19.7 & 1.756  & 19.62  & 20/11/97  &  $3\times2048$ \\
 SGP2:25    &  RQQ   & 00 52 07.60  & $-$29 17 50.2 & 1.868  & 19.55  & 28/08/97  &  $4\times2048$ \\
 SGP2:11    &  RQQ   & 00 52 38.47  & $-$28 51 12.9 & 1.976  & 19.77  & 02/09/97  &  $4\times2048$ \\
 SGP3:39    &  RQQ   & 00 55 43.41  & $-$28 24 09.7 & 1.964  & 19.65  & 06/09/97  &  $4\times2048$ \\
 SGP4:39    &  RQQ   & 00 59 08.88  & $-$27 51 24.7 & 1.716  & 19.64  & 30/08/97  &  $3\times2048$ \\
 1148+56W1  &  RLQ   & 11 50 44.8   & $+$56 32 56.0 & 1.782  & 19.69  & 30/06/97  &  $4\times2048$ \\
 PKS1524-13 &  RLQ   & 15 26 59.44  & $-$13 51 01.3 & 1.687  & 19.69  & 31/07/97  &  $4\times2048$ \\
 B2-2156+29 &  RLQ   & 21 58 42.0   & $+$29 59 08.0 & 1.759  & 19.29  & 05/11/97  &  $3\times2048$ \\
 PKS2204-20 &  RLQ   & 22 07 33.94  & $-$20 38 34.9 & 1.923  & 19.49  & 04/09/97  &  $4\times2048$ \\
 4C45.51    &  RLQ   & 23 54 22.27  & $+$45 53 05.2 & 1.992  & 19.79  & 14/11/97  &  $4\times2048$ \\
 SA107-626  &  PSF star & 15 40 05.31 & $-$00 17 29.2 & 0    & 12.77  & 17/02/98  &  $4\times256$  \\
 SA107-627  &  PSF star & 15 40 07.45 & $-$00 17 23.0 & 0    & 13.47  & 30/07/97  &  $4\times256$  \\ \hline

\end{tabular}
\end{minipage}
\end{table*}

\section{The Observations} 

Previous observations of quasar hosts with HST have demonstrated the
importance of careful sample construction, the matching of samples and
the selection of filter bandwidths to avoid the inclusion of strong
emission lines. In this section we describe the selection criteria
used in the construction of our quasar samples as well as the
considerations which shaped the final observing strategy. The three
samples resulting from these considerations, at $z\sim 0.4$, 1 and 2,
are shown in Figure~1. The observations of the $z\sim0.4$ sample,
which use WFPC2 rather than NICMOS, were incomplete at the time of
writing and discussion of the images of these objects is deferred to a
subsequent paper. Details of the two high-redshift ($z\sim 1$, 2)
samples, including redshifts, positions and observing dates, are given
in Table~1. It is these two samples which are the subject of the
present paper.

\subsection{Sample design and choice of filters}

\subsubsection{Luminosity constraints} 
We have confined our selection of high-redshift quasars to the
absolute magnitude range $-24 > M_V > -25$. This luminosity band is
comparable to that of a subset of our low-redshift ($0.1\leq z \leq 0.25$)
quasar sample observed with WFPC2 on HST (McLure et al. 1999; Dunlop
et al. 2001), allowing us to use this latter sample as a low-redshift
baseline against which to measure any redshift-dependent change in the
properties of the hosts for a narrow range of quasar
luminosities. Also, experience from low-$z$ studies, as well as the
results of simulations using artificial quasar/host combinations and
our 2-D modelling algorithm, demonstrates that in order to reliably
recover the luminosities and scalelengths of their host galaxies it is
currently desirable to limit the high-redshift samples to quasars of
such moderate-to-low optical luminosities.

\subsubsection{Filter/redshift combinations} 
Having defined the luminosity range of the sample, the next priority
was to break the flux-density/redshift correlation which inevitably
arises in samples derived from flux-limited surveys, by ensuring that
we sample a comparable range of luminosities at several different
redshifts. However, to obtain a clean view of the starlight from the
host galaxies at all redshifts it is also necessary to ensure that, as
in our previous WFPC2 study at $z\simeq 0.2$, the spectrum of the host
galaxy is always observed at $\lambda_{rest}>4000$\AA~ and yet is
uncontaminated by either [O{\sc iii}] or H$\alpha$ line emission. In
the $z\simeq0.2$ study this entailed using the F675W filter on
WFPC2. For the current study two NICMOS filters were selected: F110M
(roughly $J$-band), which samples the desired rest-frame wavelengths
for redshifts $0.83\leq z \leq 1.00$; and F165M (roughly $H$-band),
for $1.67\leq z \leq 2.01$. A third filter, F184W (roughly $I$-band)
on WFPC2, was also chosen to bridge the gap between the NICMOS
observations and the McLure et al. (1999) low-redshift study by
observing a third group of objects at $0.32\leq z \leq 0.43$. (HST
possesses slightly wider filters in all three of these wavelength
regimes, but using these would have led to contamination by [O{\sc
iii}] or H$\alpha$.) Thus the current study comprises three samples at
redshifts of $z\sim0.4$, 1 and 2, using filters which approximate to
standard $I$, $J$ and $H$-bands respectively.  At the time of writing
the WFPC2 observations of the sample at $z\sim0.4$ were
incomplete. Hence, the current paper deals only with the NICMOS
observations of the two high-$z$ quasar samples.

Since, at each redshift, our chosen filter corresponds to rest-frame
$V$-band, this observing strategy also obviates the need for
k-corrections when calculating host-galaxy luminosities.  Absolute
$V$-band magnitudes can be calculated directly from the observed $R$,
$I$, $J$ and $H$ magnitudes without the need to assume any particular
spectral shape (and hence stellar population) for the hosts.

\subsubsection{Radio loudness} 
In order to attempt a meaningful comparison of the properties of RLQ
and RQQ hosts as a function of redshift we selected five radio-loud
and five radio-quiet quasars in each redshift regime, giving a total
of 30 objects.  In order to perform a clean comparison of the hosts of
RLQs and RQQs at each epoch it was important to ensure that the RLQs
are genuinely radio-loud ($P_{5 GHz} > {\rm 10^{25} W Hz^{-1}
sr^{-1}}$), that the RQQs are genuinely radio-quiet ($P_{5 GHz} <
10^{24.5} {\rm W Hz^{-1} sr^{-1}}$), and that within each redshift bin
the optical luminosity distributions of the two types are well
matched.  This is not a trivial task. It requires that we select only
RQQs which have been observed with the VLA with high sensitivity but
have not been detected, and requires that we confine our RLQ sample to
steep-spectrum objects whose intrinsic radio luminosity is not being
artificially boosted by relativistic beaming. 

The relatively small number of optically-faint steep-spectrum RLQs,
along with our stringent definition of radio quietness (few RQQs have
been observed with sufficient sensitivity in the radio to meet our
self-imposed luminosity requirement), means that the resulting three
quasar samples at $z\sim 0.4$, 1 \& 2 (illustrated in Figure~1)
are almost uniquely defined.

\subsection{Observing strategy}

\subsubsection{Choice of detector} 
NICMOS contains three HgCdTe arrays, NIC1, NIC2 and NIC3, each
consisting of $256\times256$ pixels. Unlike CCDs, each pixel in the
array is independent of its neighbours and can be read
non-destructively, allowing multiple read-outs to be performed
throughout the duration of a single exposure. This ability can be
utilised for the recognition and removal of cosmic ray events
during the course of the exposure.

We opted to use NIC1, the smallest of the three detectors, with a
field of view $11''\times11''$ in extent. The NIC1 PSF is diffraction
limited for wavelengths $\geq 1~\mu$m and its $0.043''$ pixels
guarantee critical sampling of the PSF over the wavelength range of
interest.

\subsubsection{Integration times} 
A total of 60 orbits of HST time was allocated for this project. From
this total, 11 orbits were devoted to the WFPC2/F814W observations of
the 10 quasars in the $z\sim0.4$ sample as well as a suitable PSF
star, leaving 49 orbits for the two high-$z$ NICMOS samples.
Simulations using the 2-D modelling algorithm developed for the
analysis of our previous HST study of quasar hosts at $z\sim0.2$,
along with pre-launch predictions of the sensitivity of NICMOS through
the F110M and F165M filters indicated that in order to detect host
galaxies comparable in size and luminosity to those found at low
redshift we would require one orbit per object for the sample at
$z\sim 1$ and three to four orbits per object at $z\sim2$ (the final
allocation of orbits is shown in Table~1). Two orbits were reserved
for observations of PSF stars through each of the two NICMOS filters
(see below).

The NICMOS detectors allow several standard observing modes, taking
advantage of the non-destructive readout capabilities of the detector
arrays. We chose to use one of the pre-defined `multi-accumulate'
(MULTIACCUM) sequences, with a total of 25 readouts at specific
intervals during each integration, which combine high sensitivity with
large dynamic range and the ability to process cosmic rays and
reconstruct saturated regions of the image. The sequence used was
MIF2048 which, with an integration time of 2048 seconds, was the
longest sequence which would fit into a single orbit.

\subsubsection{Determination of the NICMOS point-spread function}
Any attempt to determine the properties of a faint, diffuse object
surrounding a much brighter point source ultimately depends on our
ability to separate the emission which is genuinely spatially extended
from light from the central source which has been artificially spread
out by the point-spread function (PSF) of the instrument.
Because of the extreme sensitivity of this process to the exact form
of the PSF it is vitally important that this is accurately known. This
urgency is compounded by the fact that the PSF is a complicated
function of the filter used and the SED of the target as well as the
position of the object within the aperture.

On-orbit verification and calibration of the NICMOS instrument did not
include the acquisition of empirical PSFs for every combination of
filter and camera. Synthetic PSFs, although normally an excellent
match in the bright central regions, are often less good at
reproducing the faint outer wings of the structure and usually fail to
account for the effects of instrumental defects and uncertainties.  We
therefore used two orbits of our allocated HST time to obtain
deep NICMOS stellar PSFs in both our chosen filters, F110M and
F165M. To safeguard against the possibility of Vega-like circumstellar
dust shells which would compromise the PSF, we observed two different
stars, SA~107-626 and SA~107-627. The stars have absolute magnitudes,
$M_{V}$, of 13.47 and 13.34 respectively, enabling us to obtain
high-dynamic range images of each star, tracing the PSF much further
into the wings than the quasar exposures, in under half an
orbit. Observations with the United Kingdom Infrared Telescope (UKIRT)
in April 1997 confirmed that the $J-K$ colours of the stars were 0.60
and 0.48 respectively - and thus a good match to the expected quasar
SED within our chosen filter bands.

\section{NICMOS data reduction}

The data were calibrated using the standard NICMOS pipeline software,
CALNICA, together with the most recent calibration files. However, certain
peculiarities of the NICMOS Camera 1 detector were not adequately
compensated for by the pipeline and had to be dealt with
separately. The most important of these were the problems of cosmic
ray persistence and the residual DC `pedestal'.

\subsection{Cosmic ray persistence} 
Since the declinations of our target quasars typically meant that they
would be visible for 3300 seconds per orbit the period of visibility
remaining after the main MIF2048 exposure and associated read time was
utilised for an additional 512-second exposure (using the pre-defined
MIF512 sequence). This exposure was placed at the beginning of the
orbit and was used to identify any `persistent' cosmic ray tracks
which had not been completely flushed from the detector before the start
of the observations. Cosmic ray persistence is a particular problem in
those observations which take place immediately after one of the
spacecraft's periodic passages through the South Atlantic Anomaly, but
can also occur at any point on HST's orbit. Unlike cosmic rays which
impact during a science exposure, the persistent tracks cannot be
removed by comparing adjacent readouts of a MULTIACCUM sequence
because they are present on the detector before the sequence
begins. These ghost images generally decay on a timescale of $\sim500$
seconds, so by placing a short exposure at the beginning of the orbit
we ensured that the problem was largely alleviated by the time the
longer science exposure began. Any remaining cosmic rays were then
identified by comparison of the MIF512 and MIF2048 exposures since,
unlike ordinary cosmic ray tracks, they appear in both images.

\subsection{DC offset} 

Of particular importance to the current study is the effect known as
the `pedestal'. This is a time- and temperature-dependent DC
offset arising from the design of the NICMOS3 array and amplifier
system. Its properties are well understood (Rieke et al. 1993), and
in principle the effects can be entirely removed as long as
contemporaneous dark frames, made at the same detector temperature as
the science exposures, are available. Where such contemporaneous darks
are not available - as in the present observations - a residual DC
offset or `pedestal' remains in the NICMOS images after the
subtraction of the dark reference file.  This offset propagates
throughout the subsequent calibration steps and leads, ultimately, to
a (small) multiple of the flat-field reference file being imprinted
onto the final image. Since the flat field contains spatial structure
on scales of $\sim1$~arcsec, similar to the expected scalelengths of
galaxies at redshifts of 1 to 2, removing the effects of the pedestal
was a priority.

The severity of the pedestal was reduced by an on-orbit correction
applied to all data taken after 12 August 1997, and by the
construction of synthetic dark reference files which were often better
able to account for the remaining offset. However, the effect is still
present to varying degrees in all of the NICMOS images in our study,
and these corrections do not apply to the three quasars in our program
which were imaged prior to 12 August 1997: SGP2:47 ($z=0.83$),
PKS0440$-$00 ($z\sim0.844$) and PKS1524$-$13 ($z\sim1.687$). The data
quality for these objects is significantly worse than for the rest of
the sample.

In order to alleviate this problem, staff at the Space Telescope
Science Institute have devised a procedure called {\it pedtherm} which
can be used to estimate and remove the DC offset in each quadrant of
an image prior to flat-fielding in CALNICA. Recalibrating our data
using this extra step led to significant reduction in the pedestal
effect, but failed to remove it entirely. 

In a final attempt to reduce the impact of the pedestal still further,
each of the calibrated NICMOS images was subjected to an additional
corrective procedure devised by ourselves. This algorithm yielded
a retrospective estimate of the height of the remaining pedestal by
insisting that the standard deviation across object-free regions of
the final flat-fielded image was minimized.  Most of the images were
noticeably improved by this final reduction step, though it proved
impossible to remove the pedestal entirely and its effects are
particularly evident in images in which the quasar is positioned close
to the boundary of two quadrants (BVF225 \& SGP2:11).

For the $z\sim 2$ quasar sample three or four separate 2048-second
$H$-band exposures were obtained for each object. Each exposure was
calibrated and treated for the pedestal effect separately before being
co-added to produce the final image.

\subsection{Other problems affecting image quality} 
Particulate contaminants (believed to be small flecks of paint) on the
surface of the NICMOS detector give rise to several regions of reduced
sensitivity, each typically a few pixels in extent. The positions of
these contaminants are accurately known, allowing the affected pixels
to be masked out. A further problem affects the upper left-hand
quadrant of the NIC1 array, a significant area of which has a
sensitivity 2-3 times lower than the mean of the detector as a
whole. The default pointing for NIC1 placed the majority of our target
quasars in the lower right-hand quadrant, thus avoiding this problem
in most cases. Only the RLQ 4C45.51 from the $z\simeq2$ sample fell
into the affected region and thus suffers from a somewhat higher noise
level than the rest of the sample.

\begin{table*}
\begin{tabular}{lccccccc}
\hline
Source & $z$ & $r_{1/2}$/kpc & $\mu_{1/2}$ & $J_{host}$ & $J_{nuc}$  & $L_{nuc}/L_{host}$ & $b/a$  \\
\hline
\multicolumn{8}{c}{Radio-Quiet Quasars}\\ 
SGP5:46 & 0.955 &  3.9  & 21.65 & 20.10 & 19.46 & 1.79 & 1.14 \\
SGP2:47 & 0.830 & -     & -     & -     & -     & -    & -    \\
BVF225  & 0.910 & 17.1  & 23.78 & 20.11 & 17.91 & 7.61 & 3.20 \\
BVF247  & 0.890 & 11.9  & 22.83 & 18.88 & 20.14 & 0.31 & 1.20 \\
BVF262  & 0.970 &  4.6  & 21.56 & 19.85 & 19.24 & 1.76 & 1.36 \\
\multicolumn{8}{c}{Radio-Loud Quasars}\\ 
PKS0440-00&0.844& 13.0  & 22.71 & 18.84 & 18.47 & 1.41 & 1.64 \\
PKS0938+18&0.943&  4.3  & 21.10 & 19.49 & 19.81 & 2.29 & 1.33 \\
3C422     &0.942& 17.0  & 22.88 & 18.29 & 17.90 & 1.43 & 1.33 \\
MC2112+172&0.878& 17.4  & 23.14 & 18.18 & 18.97 & 0.48 & 1.02 \\
4C02.54   &0.976& 10.4  & 21.62 & 19.32 & 17.61 & 4.82 & 4.08 \\
\hline
\end{tabular}
\caption{Results from the two-dimensional modelling of the
$z\simeq0.9$ quasar sample. The table gives the fits achieved using an
$r^{1/4}$ (de Vaucouleurs) model for the galaxy's surface brightness
profile, and assuming a cosmology with $H_{0}=50$~km~s$^{-1}$Mpc and
$\Omega_{m}=1.0$, $\Omega_{\Lambda}=0.0$. Column 3 lists the half-light
radius, $r_{1/2}$, of the host and column 4 the surface brightness,
$\mu_{1/2}$, at this radius ($J$ magnitudes arcsec$^{-2}$). Columns
5, 6 \& 7 list the apparent host and nuclear $J$-band magnitudes (with
associated uncertainties of $\sim 0.4$ and $\sim 0.3$ magnitudes
respectively) and the ratio of nuclear to host-galaxy luminosity,
whilst column 8 gives the axial ratio of the host.}
\end{table*}

\section{Data analysis}

In this section we describe the process by which information about the
underlying host galaxies was extracted from the NICMOS images of the
quasars. 

\subsection{The {\bf z$\simeq$1}  sample}

\begin{figure*}
\vspace{5.4cm}
\centering
\setlength{\unitlength}{1mm}
\includegraphics{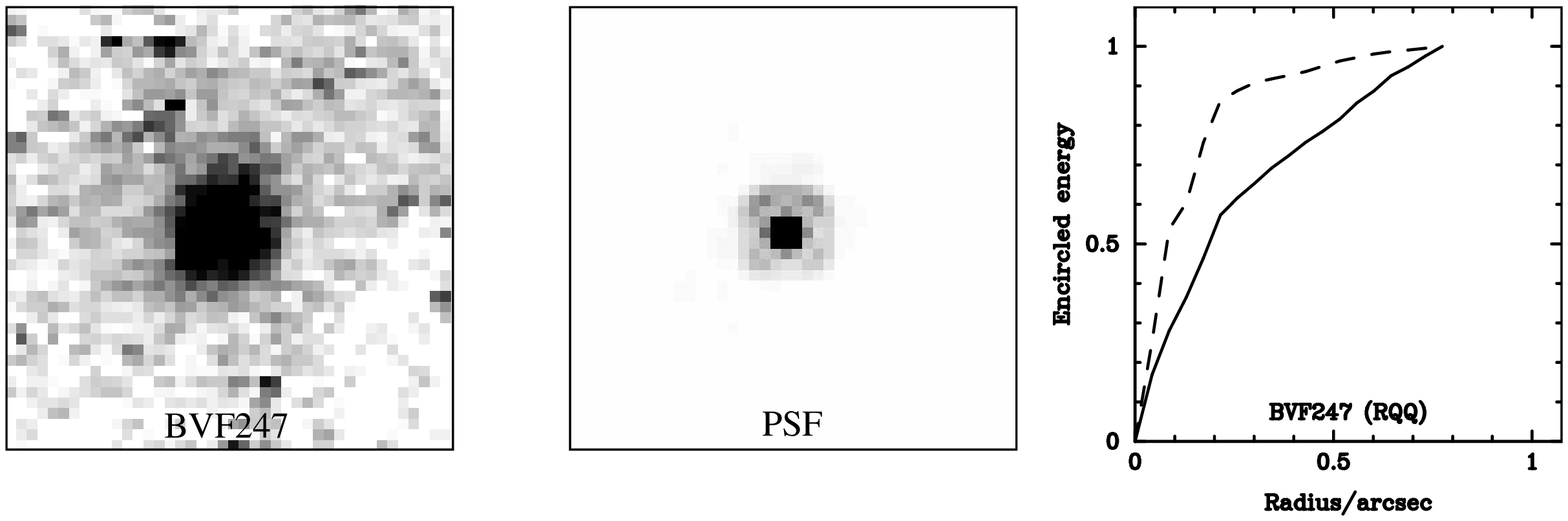}
\caption{Results of PSF subtraction on the quasar BVF247. The
left-hand panel shows the $J$-band image of the quasar after
subtraction of a stellar PSF. Prior to subtraction the PSF was
re-centred and scaled such that the value of the central pixel was
0.84 times that of the corresponding pixel in the quasar image. Using
a larger PSF:quasar ratio leads to an over-subtracted image in which
the flux ceases to rise monatonically towards the centre, causing a
characteristic dip in the values of the central few pixels.  The image
is $1''.75 \times 1''.75$ in size, with greyscale levels extending
from 0\% (white) to 25\% (black) of the maximum flux in the image. The
central panel shows the NICMOS $J$-band PSF, with an identical
greyscale cut from 0\% to 25\% of the peak. The PSF is considerably
more compact at the 25\% level than the subtracted quasar image, which
clearly contains a significant component of extended emission
underlying the quasar point source. The right-hand panel shows a
normalised encircled energy diagram for the {\it unsubtracted} quasar
image. The plot shows the cumulative flux in a circular aperture of
increasing radius (with increments of one pixel), centred on the
brightness peak and normalised at the radius at which the background
noise in the quasar image becomes significant (defined here to be the
radius beyond which the enclosed flux ceases to increase
monotonically). The solid line shows the quasar point source plus
underlying host galaxy whilst the dashed line shows the equivalent
curve for a pure (stellar) PSF. The shallower slope of the quasar
profile is further evidence of an underlying extended flux component.}
\end{figure*}

In most cases the presence of a significant component of extended
emission underlying the nuclear point source can be inferred simply
from a visual inspection of the $J$-band quasar images. The evidence
is particularly obvious between radii of 0.1 and 0.2 arcsec from the
nucleus, where the first minimum in the NICMOS PSF occurs.

\begin{figure*}
\vspace{5.5cm}
\centering
\setlength{\unitlength}{1mm}
\includegraphics{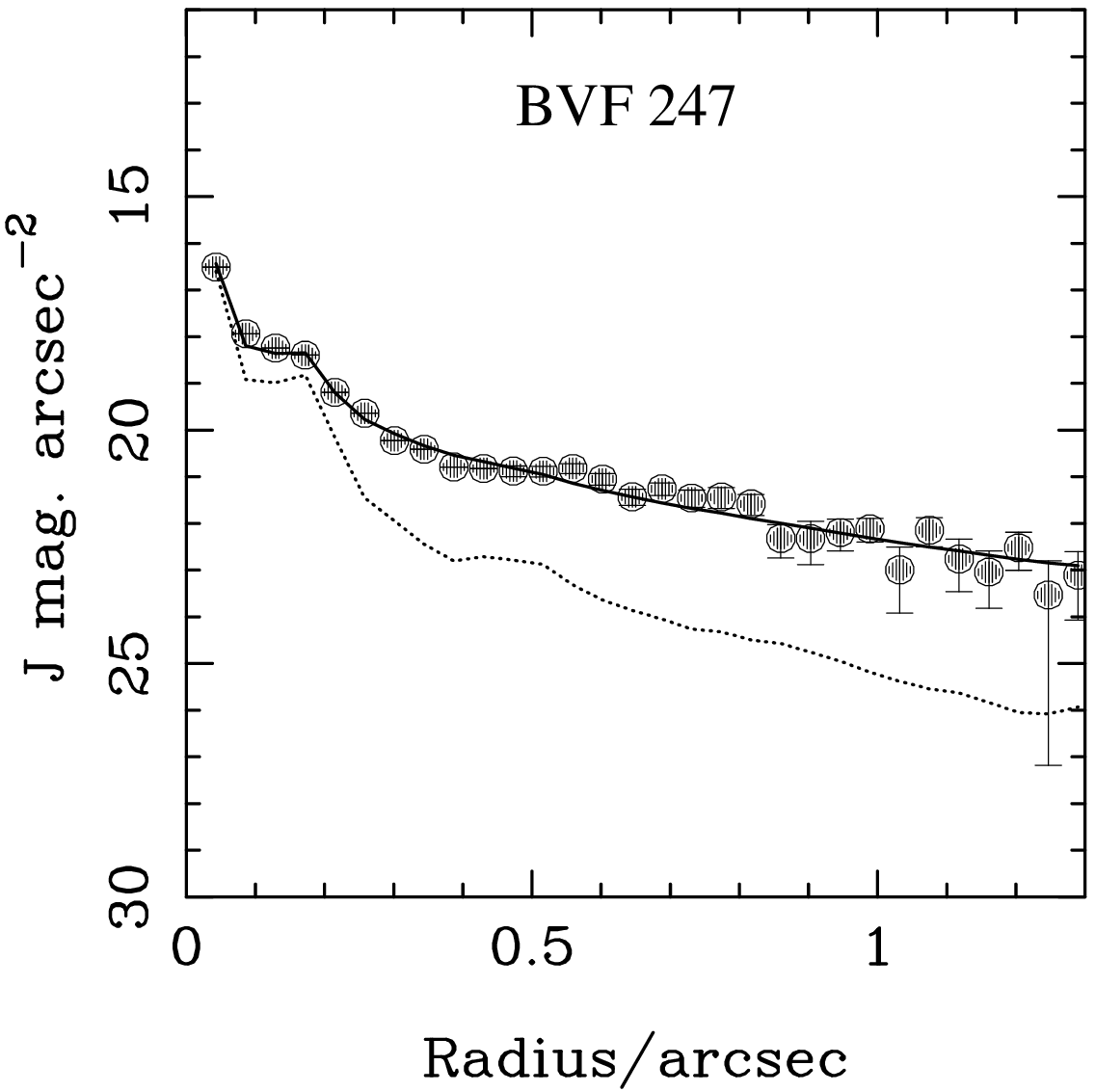}
\includegraphics{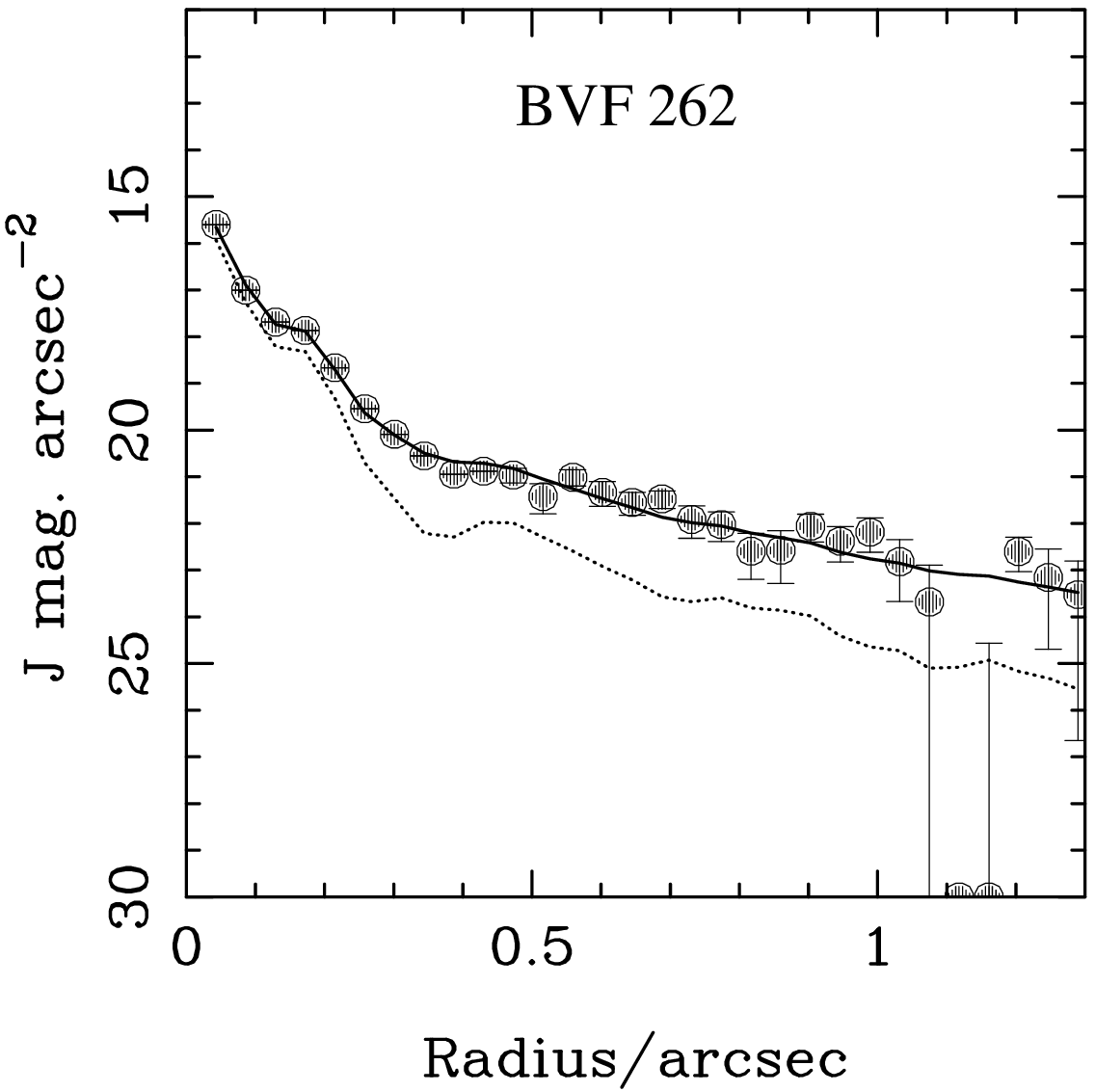}
\includegraphics{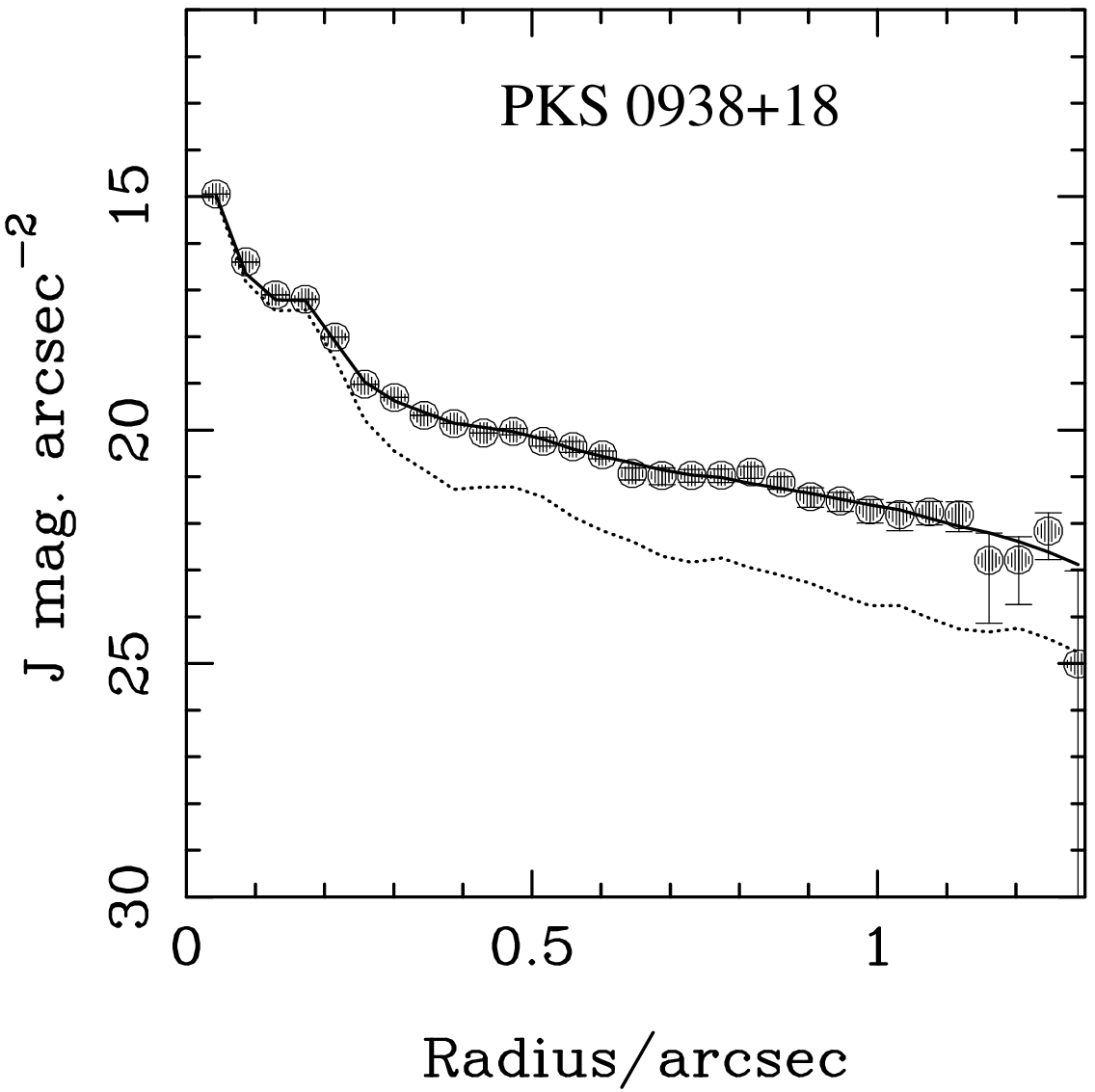}
\caption{Examples of azimuthally-averaged surface-brightness profiles
for three members of the $z\sim1$ quasar sample observed through the
F110M filter (from left to right: the RQQs BVF247 \& BVF262 and the
RLQ PKS0938+18). In each case the upper (solid) line shows the
best-fitting model (galaxy plus quasar), whilst the lower, dotted line
shows the contribution due to the nuclear (quasar) component only. In
all cases the fit was obtained by assigning a de Vaucouleurs
($r^{1/4}$) law to the host galaxy. Model parameters for each object
are listed in Table~2.}
\end{figure*}

Simple subtraction of a scaled stellar PSF provides further evidence
for the presence of an extended component in the majority of the
quasar images, and an example of a PSF-subtracted image is shown in
Figure~2.  However, in order to accurately describe the distribution
of this extended light we applied the algorithm developed for the
analysis of our previous HST/WFPC2 images of low-redshift quasars
(McLure et al. 1999; Dunlop et al. 2001). Full details of this
algorithm are given by McLure, Dunlop \& Kukula (2000). The algorithm
uses $\chi^{2}$-minimization to match a synthetic quasar$+$host to
the HST image, with nuclear luminosity, galaxy luminosity, galaxy
scalelength, axial ratio, and position angle as free parameters. The
galaxy's surface-brightness profile can be set to either an
exponentially decaying (disc) function or an $r^{1/4}$ de Vaucouleurs
law (characteristic of an elliptical galaxy) and the resulting fits
compared to determine the morphology of the underlying stellar
population. In practice, in the majority of the NICMOS images the
noise was too high to allow us to distinguish between a disc or an
elliptical profile with statistical confidence. In the case of the
three most prominent host galaxies (BVF247, BVF262 and PKS0938+18) the
preference for a de Vaucouleurs profile was more marked and we show
the results of these fits in Figure~3, but the formal significance of
this preference is only marginal.

Table~2 lists the results of 2-D modelling for all ten of the F110M
($J$-band) images, assuming an $r^{1/4}$ de Vaucouleurs
surface-brightness profile. Note that the software could not converge
on an acceptable fit for the RQQ SGP2:47, the image with the worst
residual pedestal problems. Despite the failure to unambiguously
determine the morphologies of the $z\sim1$ quasar hosts, in most other
respects these galaxies do appear to be consistent with the elliptical
hosts uncovered for comparably luminous quasars at $z \simeq 0.2$ by
our $R$-band WFPC2 HST study (McLure et al. 1999; Dunlop et
al. 2001). Half-light radii range from $\sim4$ to 20~kpc and
nuclear-to-host luminosity ratios are typically $1 - 2$ (the RQQ
BVF225 appears to have an unusually high nuclear:host ratio $\simeq 8$
but the image is unusually noisy and the resulting model fit is poorly
constrained). We note that the distribution of axial ratios of the
hosts peaks at a value of $b/a \sim 1.2$, consistent with them being
drawn from a population of elliptical galaxies (Sandage, Freeman \&
Stokes 1970; Ryden 1992).

\subsection{The {\bf z$\simeq$2} sample}

Analysis of the $H$-band (F165M) images of the $z\sim2$ quasars proved
far more problematic. One object, the RLQ 1148+56W1, was not detected
due to a combination of positional uncertainty and the HST pointing
errors. For the remaining nine quasars, despite the fact that the
nominal sky background at $H$ is little different to that at $J$, by
selecting objects of the same absolute magnitudes as those at lower
redshifts, the apparent brightness of the targets had of course
decreased substantially.

\begin{figure*}
\vspace{5.4cm}
\centering
\setlength{\unitlength}{1mm}
\includegraphics{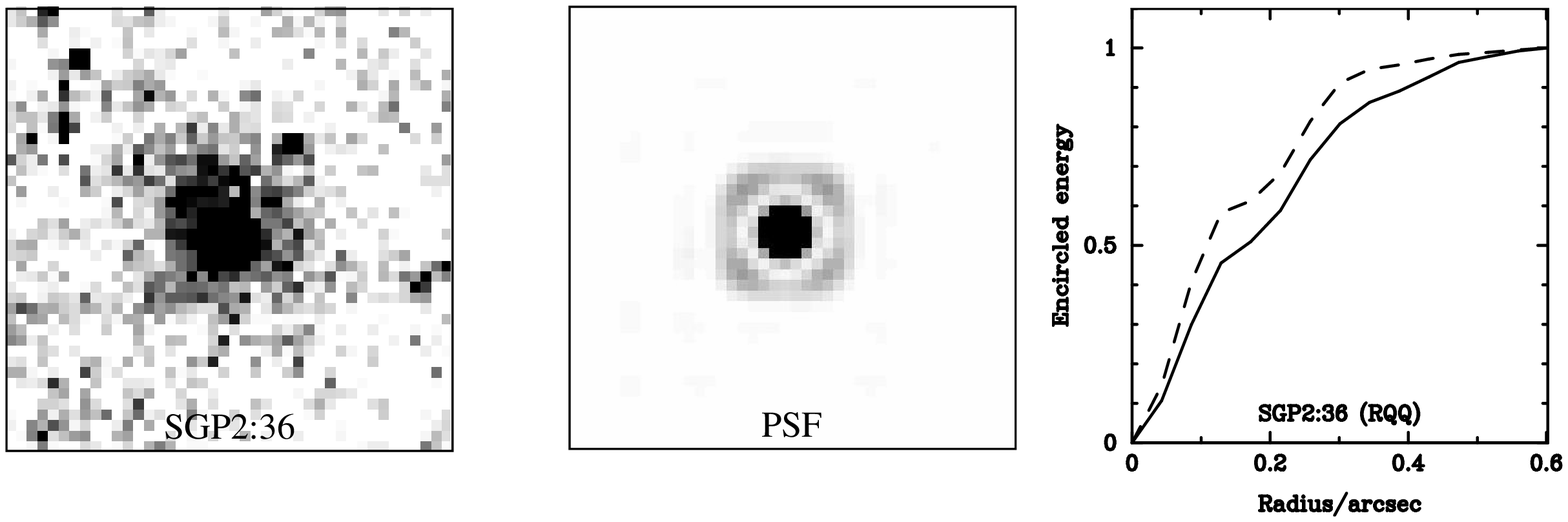}
\caption{Results of PSF subtraction on the quasar SGP2:36. The
left-hand panel shows the $H$-band image of the quasar after
subtraction of a stellar PSF. Prior to subtraction the PSF was
re-centred and scaled such that the value of the central pixel was 0.8
times that of the corresponding pixel in the quasar image. Using a
larger PSF:quasar ratio results in an over-subtracted image.  The
image is $1''.1 \times 1''.1$ in size, with greyscale levels extending
from 0\% (white) to 25\% (black) of the maximum flux in the image. The
central panel shows the NICMOS $H$-band PSF, with an identical
greyscale cut from 0\% to 25\% of the peak. Although the difference is
less marked than in the $J$-band example of Figure~2, the residual
flux in the subtracted image is more extended than the PSF and cannot
be accounted for by a point source.  The right-hand panel shows a
normalised encircled energy diagram for the {\it unsubtracted} quasar
image. The plot shows the cumulative flux in a circular aperture of
increasing radius (with increments of one pixel), centred on the
brightness peak and normalised at the radius at which the background
noise in the quasar image becomes significant (defined here to be the
radius beyond which the enclosed flux ceases to increase
monotonically). The solid line shows the quasar point source plus
underlying host galaxy whilst the dashed line shows the equivalent
curve for a pure (stellar) PSF. The slope of the quasar curve is less
steep than that for the pure PSF, providing further evidence for an
extended component of emission underlying the quasar point source.}
\end{figure*}

\begin{table*}
\begin{tabular}{lccccrc}
\hline
Source & $z$ & $r_{1/2}$/kpc & $H_{host}$ & $H_{nuc}$ & $L_{nuc}/L_{host}$ & b/a \\
\hline
\multicolumn{7}{c}{Radio-Quiet Quasars}\\ 
SGP2:11   & 1.976 & ? ($<$10)   &20.65 &18.97 &4.70 &- \\
SGP2:25   & 1.868 & ? ($<$10)   &19.89 &19.60 &1.31 &- \\
SGP2:36   & 1.756 & $\sim$5     &19.74 &19.98 &0.80 &- \\
SGP3:39   & 1.964 & ? ($<$10)   &19.77 &19.55 &1.16 &- \\
SGP4:39   & 1.716 & ? ($<$10)   &21.60 &18.86 &12.43&- \\
\multicolumn{7}{c}{Radio-Loud Quasars}\\ 
PKS1524$-$13& 1.687 & $\sim$5 &19.36 &18.10 &3.22 &- \\
B2-2156+29& 1.753 & 16.0    &17.87 &17.97 &0.91 &1.77 \\
PKS2204$-$20& 1.923 & $\sim$5 &20.66 &18.57 &6.87 &- \\
4C45.51   & 1.992 & 17.9    &17.86 &17.48 &1.42 &1.23 \\
\hline
\end{tabular}
\caption{Results from the two-dimensional modelling of the $z\simeq2$
quasar sample. The table gives the fits achieved using an $r^{1/4}$
(de Vaucouleurs) model for the galaxy's surface brightness profile,
and assuming a cosmology with $H_{0}=50$~km~s$^{-1}$Mpc and
$\Omega_{m}=1.0$, $\Omega_{\Lambda}=0.0$. Column 3 lists the half-light
radius, $r_{1/2}$, of the host galaxy for those objects in which the
model was able to converge on a definite value. For SGP2:36,
PKS1524$-$13 and PKS2204$-$20 the model showed a preference for a
small ($\sim 5$~kpc) host, but this value must be treated with
caution. For the remaining objects, no reliable estimate of the galaxy
size could be obtained, although in view of the low luminosity of the
host, they are unlikely to be large. Columns 4, 5, \& 6 list the
apparent host and nuclear $H$-band magnitudes (with estimated
uncertainties of 0.75 and 0.3 magnitudes respectively) and the ratio
of nuclear to host-galaxy luminosity, whilst column 7 gives the axial
ratio of the host, where available.}
\end{table*}

These factors, coupled with the wider $H$-band PSF, meant that clear
evidence for an extended component was immediately apparent in only
two of the images (the RQQ SGP2:36 and the RLQ B2-2156$+$29). In
several objects careful subtraction of a scaled PSF does reveal the
presence of an underlying extended component although, as the example
shown in Figure~4 demonstrates, very little useful information on the
quantity and extent of the emission can be derived from this
relatively crude procedure.

In order to carry out a more rigorous search for the presence of
underlying galaxies we performed 2-D modelling of the $H$-band images
using the algorithm described in the previous subsection.  As before,
we assumed an $r^{1/4}$ (de Vaucouleurs) surface brightness profile
for the putative galaxy and used our stellar $H$-band image to
represent the NICMOS PSF. The results of this 2-D modelling are listed
in Table~3.

For two objects, the RLQs B2-2156+29 \& 4C45.51, the model rapidly
converged on a solution, finding luminous, extended galaxies
(half-light radii $\sim 2$~arcsec). For the remaining seven objects,
in order to obtain an adequate 2-D fit to the images the algorithm
once again found it necessary to include some extended emission in
addition to the quasar PSF.  However, the algorithm experienced
difficulty in converging on a unique solution for the underlying
galaxy. The problem was identified as an inability to determine the
luminosity and scalelength of the galaxy simultaneously, due to the
fact that the amount of extended flux was generally low, and its
distribution lay well within the wings of the $H$-band point spread
function.

In order to encourage the algorithm to converge on a unique solution
we reduced the number of free parameters by using a circular galaxy
model and fixing the scalelength at either 5 or 10~kpc. By comparing
the quality of the resulting fits we could assess whether the `small'
or `large' galaxy model gave a better match.  In three cases, the RQQ
SGP2:36 and the RLQs PKS1524$-$14 \& PKS2204$-$20, this produced a
significant preference for the smaller, 5~kpc model. Note that, though
the extended emission in SGP2:36 is clearly visible in the $H$-band
image, the small extent of the galaxy relative to the wings of the
$H$-band PSF was enough to frustrate the modelling software until the
number of degrees of freedom was reduced.

For the last four objects neither model gave a significantly better
fit to the data, though both gave statistically acceptable results and
the flux of the host galaxy remained stable to within $\pm 0.75$ mags.
However, in view of the relatively small amount of extended flux, it seems
unlikely that the galaxies are much larger than the FWHM of the NICMOS PSF
(0.19~arcsec $\sim 2$~kpc). In Table~3 we therefore adopt a tentative
upper limit of 10~kpc.

\section{Discussion}

Despite the difficulty in determining morphologies and, at $z\sim2$,
scalelengths for the quasar hosts, the information obtained from the
NICMOS images allows us to compare the galaxies with other types of
active and inactive galaxies at the same redshifts and to investigate
the link between quasar and galaxy evolution. We first discuss the
implications of our relatively robust results at $z \simeq 1$, before
considering the implications of our $z \simeq 2$ results in the context
of theories of quasar/galaxy evolution.

\subsection{Comparison of RLQ and RQQ hosts at {\bf z$\simeq$1}} 

In our study of low-redshift ($z\sim0.2$) radio-loud and radio-quiet
quasars (McLure et al. 1999, Dunlop et al. 2001, Hughes et al. 2000,
Nolan et al. 2001), for quasars with $M_{V} \leq -24$ we found no
statistically significant differences between the host galaxies of the
two types of quasar. Both RLQs and RQQs seem to lie in large
($r_{1/2}\sim 10$~kpc, luminous ($L > L^{\star}$) elliptical galaxies
with colours and ages consistent with those of inactive massive
elliptical galaxies at the same redshift.

At $z\sim 1$, although we are unable to determine the morphology of
the galaxies with confidence, we find that once again the properties
of the radio-loud and radio-quiet hosts are statistically consistent
with one another, although there is some evidence of a trend towards a
larger luminosity difference than was found at $z \simeq 0.2$. The
mean $J$-band magnitude for the RLQ hosts is $18.8\pm0.3$ and that of
the RQQs is $19.7\pm0.3$, whilst the mean scalelengths are
$12.4\pm2.4$~kpc and $9.4\pm3.2$~kpc respectively. A similar tendency
towards somewhat less massive hosts for RQQs was also seen in the
$z\simeq0.2$ sample (McLure et al. 1999; Dunlop et al. 2001), although
once again the difference was not statistically significant. McLure et
al. speculate that this might indicate that the low-$z$ RQQs contain
smaller black holes than their radio-loud counterparts, but are
accreting gas with greater efficiency, to produce comparable
luminosities.  Larger samples will be required at both redshifts in
order to determine whether this apparently small but persistent
difference is in fact real.

\subsection{Comparison of quasar hosts and radio galaxies at z$\simeq$1} 

\begin{figure}
\vspace{7.7cm}
\centering
\setlength{\unitlength}{1mm}
\includegraphics{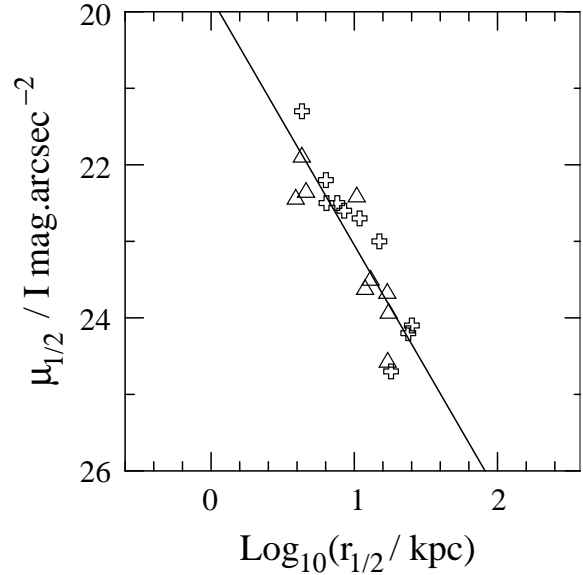}
\caption{The apparent $I$-band Kormendy relation described by the
$z\sim1$ quasars (triangles) imaged with NICMOS through the F110M
($\simeq J$) filter. Conversion to $I$ magnitudes has been carried out
assuming $I-J$ colours of 0.8, a typical value for elliptical galaxies
at $z\sim1$ (Fasano et al. 1998). Also shown are the $z\sim0.8$ 3CR
radio galaxies from the sample of Best, Longair \& R\"{o}ttgering
(1997; 1998) (crosses). These objects were imaged in $I$-band with
WFPC2 on HST and the images reanalysed by McLure \& Dunlop (2000) with
the same algorithm used to model the NICMOS images in the current
study. The solid line shows a least squares fit to the combined
samples, and has a slope of $3.23\pm0.42$ (cf 3.5 for the radio
galaxies alone).}
\end{figure}

In our previous HST study of quasar hosts at $z\sim 0.2$, as well as
radio-loud and radio-quiet quasars we also included a sample of FRII
radio galaxies which was carefully matched to the RLQ sample in terms
of extended 5~GHz luminosity and radio spectral index (Taylor et
al. 1996). This enabled us to compare the properties of the three main
types of powerful active galaxy in the local universe, and to test
models which claim to unify RLQs and radio galaxies via viewing angle
effects.  We found that the radio galaxies were indistinguishable from
the hosts of RLQs and luminous RQQs, having similar sizes, colours and
luminosities and following a Kormendy relation identical to that of
normal, inactive massive elliptical galaxies (McLure et al. 1999;
Dunlop et al. 2001).

Due to scheduling pressures it was not practical to include a
comparison sample of radio galaxies in the current NICMOS
study. However, by utilising the data in the HST archive we are able
to construct a {\it post hoc} sample of radio galaxies which is
roughly matched to the quasars at $z\sim 0.9$.

Of all the radio galaxy images in the archive the most useful in terms
of redshift, radio luminosity and observing waveband are the $I$-band
images of 3CR radio galaxies made with WFPC2 by Best, Longair \&
R\"{o}ttgering (1997; 1998). These images have been reanalysed by McLure \&
Dunlop (2000) with the same 2-D modelling software used to analyse the
current NICMOS images. By assuming an $I-J$ colour of 0.8 for
elliptical galaxies at $z\sim1$ (Fasano et al. 1998) it becomes
possible to make a comparison between the quasar hosts imaged in
$J$-band with NICMOS and the $I$-band radio galaxy images.

We find that the $I$-band properties of the quasar hosts are very
similar to those of the radio galaxies: mean scalelengths (with
standard errors) are $12.6\pm2.3$ (RGs) and $11.0\pm1.8$ (quasar
hosts), whilst the mean absolute $I$ magnitudes are $-24.62\pm0.13$
(RGs) and $-24.59\pm0.21$ (quasar hosts).

\begin{figure*}
\vspace{7.0cm}
\centering
\setlength{\unitlength}{1mm}
\includegraphics{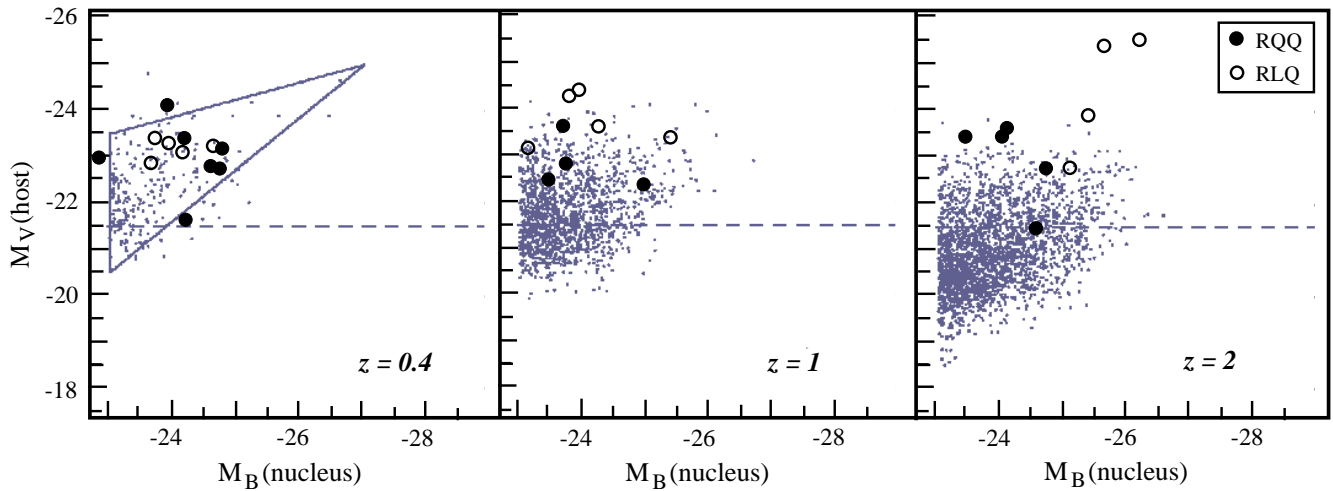}
\caption{Host galaxy versus quasar (nuclear) absolute magnitudes at
redshifts of $0.4$, $1$ and $2$. The dashed line shows the value of
$L^{\star}_{V}$ for galaxies in the local universe. The small grey
points are the simulated data from the semi-analytic model of
Kauffmann \& Haehnelt (2000), showing the predicted evolution of the
relationship between $M_{V}$ of the host and $M_{B}$ of the active
nucleus, and assuming the same cosmology used in the current
paper. The large symbols show the results of the current NICMOS study,
with filled circles representing RQQs and open circles RLQs. Since we
currently lack data for objects at $z=0.4$, in the left hand panel we
plot the subset of RLQs and RQQs with $-24\geq M_{V}(total) \geq -25$
from our WFPC2 study of quasars at $z\sim0.2$ (McLure et al. 1999;
Dunlop et al. 2001). The triangle in the left hand panel shows the
locus of the quasar sample of McLeod, Rieke \& Storrie-Lombardi
(1999). Since at each redshift our chosen filter corresponds to
rest-frame $V$-band, the absolute $V$-band magnitudes can be
calculated directly from the observed $R$, $J$ and $H$ magnitudes
without the need to assume any particular spectral shape for the
hosts. For the quasars we have converted from $M_{V}$ to $M_{B}$ using
a colour index of $B-V=0.4$ (equivalent to a spectrum of the form
$f(\nu) \propto \nu^{-0.2}$).  (Figure adapted from Figure~12 of
Kauffmann \& Haehnelt (2000).)  }
\end{figure*}

Given that the redshift distributions of the 10-object 3CR galaxy
subsample and the new $z\sim1$ quasar sample are similar, it is also
possible to investigate whether their Kormendy relations are
compatible, without the need to make surface-brightness
corrections. Once again, we can convert the NICMOS $J$ magnitudes
for the quasars to $I$-band values by assuming an $I-J$ colour of
0.8. The two sets of data are shown in Figure~5, where it can be
seen that the Kormendy relations for both quasars and radio galaxies
are statistically indistinguishable. A least-squares fit to all the
data points produces the relationship $\mu_{1/2} = 3.23_{\pm0.42} {\rm
log}r_{1/2} + 19.82_{\pm0.4}$ (for the radio galaxies alone the fit
has a slope of 3.5; McLure \& Dunlop 2000). Thus at these redshifts
both the quasar hosts and radio galaxies appear to be derived from the
same parent population.  This result is consistent with the
unification of RLQs and radio galaxies via viewing-angle effects, a
scenario which is already well supported at low redshifts.

The lack of bright nuclear point sources in the radio galaxies means
that their morphologies can be determined unambiguously; as expected
all have radial surface brightness profiles which are well described
by a de Vaucouleurs law. Thus, although our analysis of the quasar
hosts cannot distinguish between an elliptical or a disc profile with
any degree of confidence, their similarity in all other respects to
the 3CR radio galaxies provides strong circumstantial evidence that,
as at low redshifts, these quasars inhabit massive elliptical
galaxies.  If this is the case it implies that, just as at low
redshift, an important prerequisite for the presence of a powerful AGN
at $z\sim 1$ is a host galaxy with a massive spheroidal component.
Within the context of hierarchical clustering models for galaxy
formation, the similarity (but for passive stellar evolution) between
both radio galaxies and quasar hosts at $z\simeq1$ and their
counterparts at $z\simeq0.2$ is most simply explained by a
cosmological model in which the massive elliptical galaxy formation
process is essentially complete by $z \simeq 1$.

\subsection{The nature of quasar hosts at z$\simeq$ 2}

The strong cosmological evolution in comoving space density of quasars
has been known for nearly 40 years. Much work has gone into empirical
fits of the evolution (Schmidt \& Green 1983; Boyle, Shanks \&
Peterson 1988; Dunlop \& Peacock 1990; Hewett, Foltz \& Chaffee 1993;
Goldschmidt \& Miller 1998; Goldschmidt et al. 1999) but this has not
led to any physical understanding of its causes.  Some recent
theoretical work (Efstathiou \& Rees 1988; Carlberg 1990; Haehnelt \&
Rees 1993; Percival \& Miller 1999) has concentrated on explaining the
evolution as being driven by evolution in the rate of galaxy merging,
following the hypothesis that quasars display a cosmologically
short-lived burst, or bursts, of activity following such a merger.
Kauffmann \& Haehnelt (2000) have used a semi-analytic model to link
the growth of supermassive black holes with the formation and
evolution of galaxy dark matter halos in a cold dark matter (CDM)
universe.  They find that the evolution in merger rate of dark halos
alone cannot explain the large evolution in quasar space density, in
agreement with the analytic work of Percival \& Miller (1999).  They
suppose that accretion timescale also varies with cosmic epoch, and
thus they effectively add a component of luminosity evolution onto the
basic model (as also proposed by Haehnelt \& Rees 1993) which can
successfully reproduce the approximately luminosity evolution that is
observed (Boyle et al. 2000).  Within this framework, quasars of a
given luminosity are powered by progressively lower-mass black holes
with increasing redshift.  Hence if black-hole mass is indeed related
to the mass of the dark halo, as might be implied from the
correlations between black-hole mass and galaxy bulge luminosity
(Kormendy \& Richstone 1995; Magorrian et al. 1998; van der Marel
1999) and between black-hole mass and galaxy velocity dispersion
(Ferrarese \& Merritt 2000; Gebhardt et al. 2000), the Kauffmann \&
Haehnelt (2000) model predicts that quasars of a given luminosity
should, on average, be found in progressively less massive host
galaxies with increasing redshift.  In fact, provided that a
relationship between black-hole mass and host galaxy mass exists at
all redshifts, then any luminosity-evolution model will predict such
an effect, and this prediction can now be tested.

\begin{figure*}
\vspace{6.5cm}
\centering
\setlength{\unitlength}{1mm}
\includegraphics{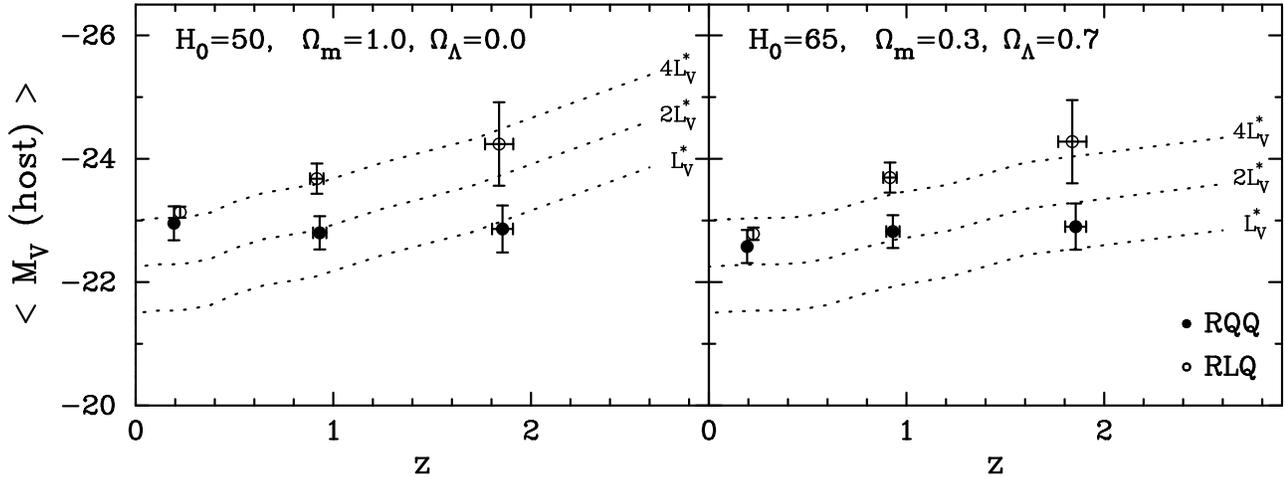}
\caption{Mean absolute $V$-band magnitude versus mean redshift for the
host galaxies of the RLQs (open circles) and RQQs (filled circles) in
the current NICMOS study. Also shown is the subset of 5 RLQs and 7
RQQs from our WFPC2 study of quasars at $z\sim0.2$ which have total
(host $+$ nuclear) luminosities in the same range as the high-redshift
samples ($-24\geq M_{V}\geq-25$). Error bars show the standard error
on the mean. The dotted lines show the luminosity evolution of present
day $L^{\star}$, $2L^{\star}$ and $4L^{\star}$ elliptical galaxies,
assuming a formation epoch of $z=5$ with a single rapid burst of
starformation followed by passive evolution thereafter. Left-hand panel:
assuming a cosmology with $H_{0}=50$~km s$^{-1}$ Mpc$^{-1}$,
$\Omega_{m}=1.0$ and $\Omega_{\Lambda}=0.0$. Right-hand panel:
$H_{0}=65$~km s$^{-1}$ Mpc$^{-1}$, $\Omega_{m}=0.3$ and
$\Omega_{\Lambda}=0.7$.}
\end{figure*}

To date, ground-based studies have not strongly supported such a
picture, finding evidence for very luminous hosts around quasars at
redshifts of $\sim 2.5$. However, these observations have concentrated
on the most luminous quasars, for which there are no counterparts at
low redshifts.  Our current HST study thus offers the prospect of the
first, proper, unbiased view of how host-galaxy luminosity varies with
redshift, in a manner which offers a direct and transparent test of
theoretical predictions.  This is not only because we have studied
quasars of similar luminosity at $z \simeq 0.2,1$ \& $2$, but also
because we have observed all the hosts at the same rest wavelength
(removing the need for k-corrections), and because host-galaxy
luminosities have been derived using an identical modelling technique
at all redshifts.

In Figure~6 we show the best-fit host and nuclear luminosities for our
quasar samples at $z \simeq 0.2$, $1$ \& $2$ superimposed onto the
scatterplot predictions for quasars/hosts at $z \simeq 0.4, 1$ \& 2
produced by Kauffmann \& Haehnelt (2000). As can be seen, the
agreement between observation and prediction is excellent for the
low-redshift sample, but becomes progressively less convincing within
the higher redshift bins as the model predictions recede, while our
derived host luminosities remain either roughly constant (in the case
of the RQQs) or actually increase with redshift (in the case of the
RLQs).  We note that the behaviour of the RLQ hosts cannot be
explained away simply as the effect of the two unusually luminous RLQs
in the $z\simeq2$ sample. (The $V$ magnitudes quoted in the
V\'{e}ron-Cetty \& V\'{e}ron catalogue are often merely extrapolations
based on data in other wavebands, so it is perhaps unsurprising that
our sample should turn out to contain two objects slightly more
luminous than our nominal upper limit of $M_{V}(total)=-25$.) Even
when these two objects are excluded, the remaining RLQ hosts at
$z\simeq 2$ show no evidence for a drop in luminosity with redshift,
and a tendency towards increasing luminosity is clearly already
present at $z\simeq1$.

To clarify the trend of host luminosity with redshift seen in our
sample, we plot mean $M_{V}$ versus redshift in Figure~7, for two
alternative cosmologies, overlaid with the predictions for a passively
evolving galaxy formed in a short-lived burst at high-redshift ($z =
5$). Passive evolution commencing at high redshift appears to be the
most reasonable starformation history to adopt, given the evidence
from Nolan et al. (2001) \& de Vries et al. (2000) that low-$z$ quasar
hosts and radio galaxies are dominated by stellar populations of age
$\simeq 12$~Gyr.

These plots demonstrate a number of potentially important
points. First, it is clear that in either cosmology the hosts of RLQs
brighten with redshift in a manner which is perfectly consistent with
pure passive evolution. Basically, the typical host of an RLQ appears
to be a passively evolving (present-day) $4L^{\star}$ elliptical. This
result agrees well with what has been found for high-redshift radio
galaxies. Specifically, if we compare our $z\simeq 2$ results directly
with $H$-band observations of radio galaxies at comparable
redshift, the average $H$-band magnitude for radio galaxies in the
redshift band $1.7 - 2.1$ discussed by Lacy, Bunker \& Ridgway (2000)
is $19.2 \pm 0.3$, while the average $H$-band magnitude for the RLQ
hosts in our $z \simeq 1.9$ sample is $18.9 \pm 0.7$.  Our results for
RLQ hosts are also consistent with the findings of a recent
ground-based imaging study of three RLQs at $z \sim 1.5$ by Falomo et
al. (2001).  Taken together, these results therefore provide further
support for a picture in which the formation of ellipticals with
sufficiently massive black holes to support radio activity (see McLure
et al. 1999), is essentially complete before $z \simeq 2$.

The trend displayed by the hosts of RQQs does appear to be somewhat
different however. While the error bars have also grown, it seems
that the gap between the luminosities of the hosts of RQQs and RLQs has
grown still further to $\simeq 1.5$ magnitudes, simply because the
typical RQQ host in our sample appears to have the same
absolute magnitude at all redshifts.  

The interpretation of this observation is dependent on choice of
cosmology.  In the flat matter-dominated cosmology, the RQQ hosts are
inconsistent with having had a constant mass in stars which has
evolved passively since $z = 5$.  Evaluation of any change in mass
then depends on the choice of star-formation history, but is a factor
of 4 under the assumption of passive stellar evolution between $z
\simeq 2$ and the present day. In the $\Lambda$-dominated cosmology,
the RQQ hosts are in fact consistent with no change in mass.  Thus,
even accounting for the apparent difference between RLQ and RQQ hosts
at $z \sim 2$, the large variation in typical host mass with redshift
predicted by Kauffmann \& Haehnelt (2000) is not observed.

Recent results from other studies of the hosts of high-redshift RQQs
also indicate that they are less luminous than their radio-loud
counterparts. In particular, Ridgway et al. (2001) report results from
NICMOS imaging of 5 RQQs at $z \simeq 2 - 3$, in which they find host
galaxies with typical luminosities of $\sim L^{\star}$. In fact two of
their RQQs are rather fainter than those discussed here, and if one
confines attention to the three objects in their sample with $M_B
\simeq -24$ their PSF-subtracted host luminosities are $\simeq 2
L^{\star}$. Any subsequent correction for PSF over-subtraction would
only raise these values further, guaranteeing rather good agreement
with the results shown in Figure~7.

Other preliminary NICMOS-based results on the luminosities of RQQ
hosts at $z \simeq 2$ have been presented by Rix et al. (1999), as an
interesting by-product of the CfA-Arizona-Gravitional-Lens-Survey
(CASTLES).  This study attempts to measure the host luminosities of
RQQ in strongly lensed systems, by correcting the observed extended
emission for the effects of the gravitational lensing, and it has been
claimed that their results mirror exactly the predictions of Kauffmann
\& Haehnelt (2000). Such a conclusion may be premature since it is
hard to assess the surface brightness biases inherent in this method,
but nonetheless it seems unlikely that the derived RQQ host
luminosities could be sufficiently underestimated (by $\sim2$ mags) to
be consistent with the hosts of RLQs and radio galaxies.

It thus seems hard to avoid the conclusion that there exists a real
difference of $\simeq 1$ mag. between the hosts of radio-loud and
radio-quiet AGN of comparable nuclear optical luminosity at $z \simeq
2$.

\subsection{Implications}

These results raise two fundamental questions. First, what are the
implications of the observed rather modest drop in RQQ host mass as a
function of redshift? Second, why is a comparable drop in host mass not
apparent for the RLQs?

The answer to the first question is that, provided the relationship
between black-hole and host-spheroid mass is basically unchanged out
to $z \simeq 2$, our results exclude any model of quasar evolution
which involves a substantial component of luminosity evolution. This
is because the rather modest drop in RQQ host mass with increasing
redshift implies that a typical RQQ black-hole power source at $z
\simeq 2$ is only radiating $\simeq 2-3$ times more efficiently than
at low-redshift. This is obviously completely at odds with pure
luminosity evolution models which, to explain the evolution of the
quasar optical luminosity function, require black holes to radiate
$\simeq 30$ times more efficiently at $z \simeq 2$. However, it is
also in conflict with more topical and apparently more realistic
models, such as the hierarchical models of Kauffmann \& Haehnelt
(2000), which still require a substantial (order-of-magnitude)
component of luminosity evolution to be invoked to explain the rapid
evolution of the quasar luminosity function (e.g. Boyle et al. 2000).

Our results are much more consistent with a picture in which the
increased availability of fuel (or, equivalently, increased frequency
of mergers) at $z \simeq 2$ results in a substantial increase in the
number density of active black holes, along with a moderate increase
in the fueling efficiency of a typical observed quasar. An increase of
a factor of 2-3 in typical fueling efficiency between $z \simeq 0.2$
and $z \simeq 2$ is certainly permitted by the results of detailed
studies of black-hole masses in nearby RQQs (Dunlop et al. 2001;
McLure et al. 2001), which suggest that on average RQQs at $z \simeq
0.2$ are radiating at only $\simeq 20$\% of their Eddington limit. It
is thus not unreasonable to find that, in an era of greater fuel
availability, our sample of $z \simeq 2$ RQQs, matched in nuclear
optical output to our low-redshift sample, could be largely produced
by black-holes (and hence hosted by spheroids) with a characteristic
mass a few times smaller than in the comparison $z \simeq 0.2$
sample. Importantly, however, an observed trend towards moderately
higher fueling efficiences with increasing $z$ does not require one to
invoke some redshift-dependent change in quasar black-hole accretion
mechanism.  On the contrary, the observed modest mass reduction is in
fact predicted by new models which attempt to explain observed quasar
evolution as a consequence of varying quasar birthrate, combined with
declining light-curves for individual quasars (Miller, Percival \&
Lambert, in preparation). In these models, the increased birth-rate at
$z \simeq 2$ leads to a statistical bias such that quasars of a given
absolute magnitude are more likely to be observed earlier in
(i.e. closer to the peak of) their declining light-curves, and
therefore closer to maximum luminosity than at low redshift, when
quasar birth is relatively rare.
 
In answer to the second question, one possible explanation is the
effect of the joint selection criteria of high radio luminosity and
moderate optical luminosity used in the definition of our RLQ
sub-samples. There is now growing evidence that luminous radio sources
are only produced by a very massive subset of the black-hole
population which powers quasars in general. In particular, the results
of Dunlop et al. (2001) and McLure \& Dunlop (2001) indicate that the
RLQs in our $z \simeq 0.2$ sample are powered by black-holes which are
typically 3 times more massive than the power sources of RQQs, and are
confined to the mass range $M > 10^9 {\rm M_{\odot}}$. If this remains
true at high redshift, and if the black-hole spheroid relationship
remains basically unchanged, then radio-based selection will
effectively guarantee host galaxies of similar mass at all
redshifts. If, as we have done in this study, one also insists that
the optical luminosities of the RLQs under study are comparable at $z
\simeq 2$ and $z \simeq 0.2$, then it would not be unexpected that
such subsamples of objects would show no evidence for the
redshift-dependent change in fueling efficiency found for the RQQs.
This does not of course mean that RLQs would not, on average, be more
efficiently fueled at high redshift, but rather that the selection
criteria used here would mitigate against us observing such an effect
in our sample.

\section{Summary}

We have presented the results of the first, major observational study
designed to determine the properties of the hosts of both radio-loud
and radio-quiet quasars from $z \simeq 2$ to the present day in a
genuinely unbiased manner. The key features of this study are: (i)
sufficient HST-based angular resolution to allow a meaningful attempt
at determining galaxy scalelengths at all redshifts; (ii) the use of
quasar samples with the same characteristic absolute magnitude at all
redshifts; (iii) our insistence that the radio-quiet quasars selected
for study are known to lie below a definite radio-luminosity
threshold; (iv) filter selection, which when coupled with careful
sample redshift constraints, guarantees line-free imaging at the same
rest-wavelength ($\simeq 5800$\AA) (removing concerns about
emission-line contamination, and obviating the need for
k-corrections); (v) the extraction of host-galaxy parameters using an
identical modelling approach at all redshifts, minimising potential
surface-brightness bias and concerns over aperture corrections; (vi)
the use of properly-sampled, high-dynamic-range PSFs derived from
observations of stars through the same filters, and on the same
regions of the relevant detectors, as the quasar images.

At $z \simeq 1$ we have been able to determine host-galaxy parameters
with sufficient accuracy to demonstrate that the hosts of both RQQs
and RLQs lie on the same Kormendy relation as deduced for 3CR radio
galaxies at comparable redshift by McLure \& Dunlop (2000). The hosts
of both RLQs and RQQs seem to have changed little, if at all, in terms
of size between $z \simeq 1$ and $z \simeq 0.2$. The typical
luminosity of the RLQ hosts has increased between $z \simeq 0.2$ and
$z \simeq 1$ by an amount that is perfectly consistent with pure
passive evolution of a mature stellar population. The typical
luminosity of the RQQ hosts is less enhanced, and indeed appears
basically unchanged. There is thus a suspicion that the host masses of
RQQs of comparable nuclear output are less massive at $z \simeq 1$,
but the significance of this result is marginal. Within a
$\Lambda$-dominated cosmology our results are certainly consistent
with the host masses of both class of quasar being unchanged between
$z \simeq 1$ and $z \simeq 0.2$.

At $z \simeq 2$, although the hosts are harder to detect and
consequently to model, we have found that host-galaxy magnitudes can
be extracted with sufficient confidence to demonstrate that the hosts
of RLQs continue to brighten as expected under the hypothesis of pure
passive evolution of hosts with non-evolving mass.  In contrast, the
hosts of RQQs again seem little changed in luminosity, and
systematically smaller in size compared to at $z \simeq 1$. This
apparently growing radio-loud/radio-quiet host-mass gap is mirrored in
other studies of radio galaxies and radio-quiet quasars at comparable
redshift, lending further support to its reality.  Even after
allowance for passive evolution, the inferred drop in RQQ host mass is
relatively modest compared to the predictions of an order of magnitude
drop in mass made by Kauffmann \& Haehnelt (2000).  The mass drop is a
factor of 4 in an Einstein-de-Sitter cosmology, assuming a passive
evolution model, and a factor of 2 if $\Omega_m = 0.3$ and $\Omega_v =
0.7$.


The lack of strong evolution in host mass with redshift argues against
models of quasar evolution, such as the latter, which invoke some
relationship between host mass and quasar luminosity together with
substantial luminosity evolution of the quasar component.  Rather,
this study indicates that the increased availability of fuel at $z
\simeq 2$ results in a substantial increase in the number density of
active black holes, along with a moderate increase in the fueling
efficiency of a typical observed quasar. This type of moderate
efficiency increase is certainly permitted by studies of the fueling
efficiency of low-redshift quasars (Dunlop et al. 2001; McLure \&
Dunlop 2001), and indeed arises naturally as a statistical effect
predicted by new models which attempt to explain observed quasar
evolution as a consequence of varying quasar birthrate, combined with
declining light-curves for individual quasars (Miller, Percival \&
Lambert, in preparation).

\section*{Acknowledgments}
The authors would like to thank the referee, D. Hines, for many useful
comments and suggestions, and E. Bergeron for help with {\it
pedtherm}. MJK, RJM \& WJP acknowledge PPARC funding.  JSD
acknowledges the enhanced research time afforded by the award of a
PPARC Senior Fellowship.  Support for this work was provided by NASA
through grant numbers GO-06776.01-95A \& GO-07447.01-96A from the
Space Telescope Science Institute, which is operated by the
Association of Universities for Research in Astronomy, Inc., under
NASA contract NAS5-26555.  Based on observations with the NASA/ESA
{\it Hubble Space Telescope}, obtained at the Space Telescope Science
Institute.  This research has made use of the NASA/IPAC Extragalactic
Database (NED) which is operated by the Jet Propulsion Laboratory,
California Institute of Technology, under contract with NASA.

\bsp

\label{lastpage}


\begin{thebibliography}{99}

\bibitem{1} Aretxaga I., Boyle B. J., Terlevich R. J., 1995, MNRAS, 275, L27
\bibitem{2} Aretxaga I., Terlevich R. J., Boyle B. J., 1998, MNRAS, 296, 643
\bibitem{3} Bahcall J. N., Kirhakos S., Schneider D. P., 1994, ApJ, 435, L11
\bibitem{4} Bahcall J. N., Kirhakos S., Schneider D. P., 1995a, ApJ, 447, L1
\bibitem{5} Bahcall J. N., Kirhakos S., Schneider D. P., 1995b, ApJ, 450, 486
\bibitem{6} Bahcall J. N., Kirhakos S., Schneider D. P., 1996, ApJ, 457, 557
\bibitem{7} Bahcall J. N., Kirhakos S., Saxe D. H., Scheider D. P.,
1997, ApJ 479, 642
\bibitem{8} Best P. N., Longair M. S., R\"{o}ttgering H. J. A., 1997, MNRAS, 292, 758
\bibitem{9} Best P. N., Longair M. S., R\"{o}ttgering H. J. A., 1998, MNRAS, 295, 549
\bibitem{10} Boyce P. J., et al., 1998, MNRAS, 298, 121
\bibitem{11} Boyle B. J., Shanks T., Peterson B. A., 1988, MNRAS, 235, 935
\bibitem{12} Boyle B. J., Fong R., Shanks T., Peterson B. A., 1990, MNRAS, 243, 1
\bibitem{13} Boyle B. J., Shanks T., Croom S. M., Smith R. J., Miller L., Loaring N., Heymans C., 2000, MNRAS, 317, 1014
\bibitem{14} Carballo R., S\'{a}nchez S. F., Gonz\'{a}lez-Serrano J. I., Benn C. R., Vigotti M., 1998, AJ, 115, 1234
\bibitem{15} Carlberg R., 1990, ApJ, 350, 505
\bibitem{16} de Vries W. H., O'Dea C. P., Barthel P. D., Fanti C., Fanti R., Lehnert M. D., 2000, AJ, 120, 2300 
\bibitem{17} Disney M. J., et al. 1995, Nature, 376, 150
\bibitem{18} Dunlop J. S., Taylor G. L., Hughes D. H., Robson E. I., 1993, MNRAS, 264, 455
\bibitem{19} Dunlop J. S., McLure R. J., Kukula M. J., Baum S. A.,
O'Dea C. P., Hughes D. H., 2001, MNRAS submitted
\bibitem{20} Dunlop J. S., Peacock J. A., 1990, MNRAS, 247, 19
\bibitem{21} Efstathiou G., Rees M. J., 1988, MNRAS, 230, 5P
\bibitem{22} Falomo R., Kotilainen J., Treves A., 2001, ApJ, 547, 124 
\bibitem{23} Fasano G., Cristiani S., Arnouts S., Filippi M., 1998, AJ, 115, 1400
\bibitem{24} Ferrarese L., Merritt D., 2000, ApJ, 539, L9
\bibitem{25} Gebhardt K., Bender R., Bower G., Dressler A., Faber S. M.,
Filippenko A. V., Green R., Grillmair C., Ho L. C., Kormendy J.,
Lauer T. R., Magorrian J., Pinkney J., Richstone D., Tremaine S.,
2000, ApJ, 539, L13 
\bibitem{26} Goldschmidt P., Miller L., 1998, MNRAS, 293, 107
\bibitem{27} Goldschmidt P., Kukula M. J., Miller L., Dunlop J. S., 1999, ApJ, 511, 612
\bibitem{28} Hamilton T. S., Casertano S., Turnshek D. A., 2001, (astro-ph/0011255)
\bibitem{29} Heckman T. M., Lehnert M. D., van Breugel W., Miley G. K., 1991, ApJ, 370, 78
\bibitem{30} Hewett P. C., Foltz C. B., Chaffee F. H., 1993, ApJ, 406, 43
\bibitem{31} Hintzen P., Romanishin W., Valdes F., 1991, ApJ, 366, 7 
\bibitem{32} Hooper E. J., Impey C. D., Foltz C. B., 1997, ApJ, 480, L95
\bibitem{33} Hughes D. H. et al., 1998, Nature, 394, 241 
\bibitem{34} Hughes D. H., Kukula M. J., Dunlop J. S., Boroson T., 2000, MNRAS, 316, 204
\bibitem{35} Hutchings J. B., 1995, AJ, 110, 994
\bibitem{36} Hutchings J. B., Morris S. C., 1995, AJ, 109, 1541
\bibitem{37} Kauffmann G., Haehnelt M., 2000, MNRAS, 311, 576
\bibitem{38} Kormendy J., Richstone D., 1995, ARA\&A, 33, 581
\bibitem{39} Lacy M., Bunker A. J., Ridgway S. E., 2000, AJ, 120, 68
\bibitem{40} Laor A., 2000, ApJ, 543, L111 
\bibitem{41} Lehnert M. D., Heckman T., Chambers K. C., Miley G. K.,
1992, ApJ, 393, 68
\bibitem{42} Lowenthal J. D., Heckman T. M., Lehnert M. D., Elias
J. H., 1995, ApJ, 439, 588
\bibitem{43} Madau P., Fergusson H. C., Dickinson M., Giavalisco M., Steidel C. C., Fruchter A. S., 1996, MNRAS, 283 1388
\bibitem{44} Magorrian J., et al., 1998, AJ, 115, 2285
\bibitem{45} M\'{a}rquez I., Petitjean P., Th\'{e}odore B., Bremer M.,
Monnet G., Beuzit J. -L., 2001, A\&A submitted, astro-ph/0103232
\bibitem{46} Marshall H. L., Huchra J. P., Tananbaum H., Avni Y., Braccesi A., Zitelli V., Zamorani G., 1984, ApJ, 283, 50
\bibitem{47} McLeod K. K., Rieke G. H., 1994a, ApJ, 431, 137
\bibitem{48} McLeod K. K., Rieke G. H., 1994b, ApJ, 420, 58
\bibitem{49} McLeod K. K., Rieke G. H., 1995, ApJ, 441, 96
\bibitem{50} McLeod K. K., Rieke G. H., Storrie-Lombardi L. J., 1999,
ApJ, 511, L67
\bibitem{51} McLure R. J., Kukula M. J., Dunlop J. S., Baum S. A.,
O'Dea C. P., Hughes D. H., 1999, MNRAS, 308, 377
\bibitem{52} McLure R. J., Dunlop J. S., 2000, MNRAS, 317, 249 
\bibitem{53} McLure R. J., Dunlop J. S., 2001, MNRAS, in press (astro-ph/0009406)
\bibitem{54} McLure R. J., Dunlop J. S., Kukula M. J., 2000, MNRAS, 318, 693 
\bibitem{56} Nolan L. A., Dunlop J. S., Kukula M. J., Hughes D. H., Boroson T., Jimenez R., 2001, MNRAS, 323, 308
\bibitem{57} Percival W. J., Miller L., 1999, MNRAS, 309, 823
\bibitem{58} Percival W. J., Miller L., McLure R. J., Dunlop J. S., 2001, MNRAS, 322, 843 
\bibitem{59} Ridgway S. E., Heckman T. M., Calzetti D., Lehnert M., 2001, ApJ, 550, 122
\bibitem{60} Rieke M. J., Winters G. S., Cadien J., Rasche R., 1993,
Proc. SPIE, 1946, 214
\bibitem{61} Rix H. -W., Falco E., Impey C., Kochanek C., Lehar J., McLeod B., Mu\~{n}oz J., Peng C., 1999 (astro-ph/9910190)
\bibitem{62} R\"{o}nnback J., van Groningen E., Wanders I., \"{O}rndahl E.,
1996, MNRAS, 283, 282
\bibitem{63} Ryden S., 1992, ApJ, 396, 445
\bibitem{64} Sandage A. R., Freeman K. C., Stokes N. R., 1970, ApJ, 268, 831
\bibitem{65} Schade D., Boyle B. J., Letawsky M., 2000, MNRAS, 315, 498
\bibitem{66} Schmidt M., Green R. F., 1983, ApJ, 269, 352
\bibitem{67} Steidel C. C., Adelberger K. L., Giavalisco M., Dickinson M., Pettini M., 1999, ApJ, 519, 1
\bibitem{68} Stockton A., MacKenty J. W., 1987, ApJ, 316, 584
\bibitem{69} Tadhunter C. N., Scarrott S. M., Draper P., Rolph C.,
1992, MNRAS, 256, 53P
\bibitem{70} Taylor G. L., Dunlop J. S., Hughes D. H., Robson E. I.,
1996, MNRAS, 283, 930
\bibitem{71} van der Marel R. P., 1999, AJ, 117, 744
\bibitem{72} V\'{e}ron-Cetty M. -P., Woltjer L., 1990, A\&A, 236, 69
\bibitem{73} V\'{e}ron-Cetty M. -P., V\'{e}ron P., 1993, ESO Sci. Rep. No. 13
\bibitem{74} Warren S. J., Hewett P. C., Osmer P. S., 1994, ApJ, 421, 412




\end{thebibliography}
\end{document}